\documentclass[3p,preview,authoryear]{elsarticle}
 
\usepackage{booktabs}           
\usepackage{tikz}               
\usepackage{mathtools}               

\usepackage{lineno,hyperref}
\usepackage[symbol]{footmisc}

\usepackage{sectsty}
\usepackage{fullpage}
\usepackage{graphicx}
\usepackage{caption}
\usepackage{epstopdf}
\usepackage{epsfig}
\usepackage{pdfpages}
\usepackage{float}
\usepackage{amsmath,bm}
\usepackage{textcomp}
\usepackage{upgreek}
\usepackage{multirow}
\usepackage{textcomp}
\usepackage{accents}
\usepackage{amsfonts}
\usepackage{lscape}
\usepackage{comment}

\modulolinenumbers[1]

\journal{arXiv}

\begin{document}

\begin{frontmatter}

\title{Transient hygro- and hydro-expansion of freely and restrained dried paper: the fiber-network coupling}

\author[1]{N.H. Vonk} \author[2,3]{W.P.C. van Spreuwel}  \author[3]{T. Anijs} \author[1]{R.H.J. Peerlings} \author[1]{M.G.D. Geers} \author[1]{J.P.M. Hoefnagels*} 

\cortext[mycorrespondingauthor]{J.P.M. Hoefnagels}
\ead{J.P.M.Hoefnagels@tue.nl}

\address[1]{Department of Mechanical Engineering, Eindhoven University of Technology, Eindhoven, The Netherlands}
\address[2]{Department of Physics, Eindhoven University of Technology, Eindhoven, The Netherlands}
\address[3]{Research \& Development, Canon Production Printing, Venlo, The Netherlands}

\begin{abstract}
The transient dimensional changes during \textit{hygro}-expansion and \textit{hydro}-expansion of freely and restrained dried, softwood and hardwood sheets and fibers is monitored, to unravel the governing micro-mechanisms occurring during gradual water saturation. The response of individual fibers is measured using a full-field global digital height correlation method, which has been extended to monitor the transient \textit{hydro}-expansion of fibers from dry to fully saturated. The \textit{hygro}- and \textit{hydro}-expansion is larger for freely versus restrained dried and softwood versus hardwood handsheets. The transient sheet-scale \textit{hydro}-expansion reveals a sudden strain and moisture content step. It is postulated that the driving mechanism is the moisture-induced softening of the so-called "dislocated regions" in the fiber's cellulose micro-fibrils, unlocking further fiber swelling. The strain step is negligible for restrained dried handsheets, which is attributed to the "dislocated cellulose regions" being locked in their stretched configuration during restrained drying, which is supported by the single fiber \textit{hydro}-expansion measurements. Finally, an inter-fiber bond model is exploited and adapted to predict the sheet-scale \textit{hygro}-expansion from the fiber level characteristics. The model correctly predicts the qualitative differences between freely versus restrained dried and softwood versus hardwood handsheets, yet, its simplified geometry does not allow for more quantitative predictions of the sheet-scale \textit{hydro}-expansion. 
\end{abstract}


\end{frontmatter}

\newpage
\section{Introduction}
During inkjet printing, liquid water is supplied on a sheet, resulting in swelling of the paper. In-proper control of swelling and shrinkage can result in unwanted (irregular) out-of-plane deformations, including cockling, fluting, waviness and curl \citep{kulachenko2005tension}. To better understand these deformations, and ultimately reduce them, the dimensional stability of paper has been intensively studied. Traditionally, the dimensional stability is studied as \textit{hygro}-expansion \citep{salmen1985plane, niskanen1997dynamic, fellers2007interaction}, during which the dimensional change of a paper sheet due to relative humidity (RH) variations is measured. However, during inkjet printing, liquid water is introduced to the material, resulting in \textit{hydro}-expansion of the paper, which, as shown in \citep{larsson2009influence, larsson2009novel}, results in significantly different dimensional changes. \\ \indent
Measurement of the \textit{hygro}-expansion is typically preferred over \textit{hydro}-expansion because in the former, the moisture content ($MC$) can be varied in a (slow) gradual manner, whereas for \textit{hydro}-expansion achieving homogeneous wetting of the sheet is challenging \citep{larsson2009influence, larsson2009novel}. During \textit{hygro}-expansion measurements, no excess water is present, enabling better fiber and sheet surface characterization and full-field monitoring. Moreover, the geometry is better preserved during \textit{hygro}-expansion, as inter-fiber bonds remain stable and out-of-plane deformations are less likely to occur due a more homogeneous $MC$ distribution enabling easier quantification of the in-plane expansion. Hence, \textit{hygro}-expansion is usually applied for studying the (continuous) transient dimensional change of paper sheets \citep{salmen1985plane, uesaka1992characterization, niskanen1997dynamic, fellers2007interaction, larsson2008influence, vonk2023frc}, and its micro-constituents, including fibers \citep{meylan1972influence, joffre2016method, vonk2021full}. In contrast, the time-dependent spectrum of the transient \textit{hydro}-expansion of fibers has not yet been reported in the literature. On the one hand, all (transverse) fiber shrinkage measurements were performed from wet to dry in one or a few discrete steps \citep{page1963transverse, tydeman1966transverse, weise1995changes}. On the other hand, \cite{larsson2009influence, larsson2009novel} did achieve small continuous $MC$ changes (similar to traditional \textit{hygro}-expansion experiments) for handsheets (using water spraying techniques). However, in contrast to \textit{hygro}-expansion experiments, introducing liquid water to a paper sheet can quickly maximize the $MC$, similar to the one-step fiber \textit{hydro}-expansion tests in \citep{page1963transverse, tydeman1966transverse, weise1995changes}. It is thought that characterization of the continuous transient \textit{hydro}-expansion behavior of individual fibers while maximizing the $MC$, up to full water immersion, may reveal mechanisms at high $MC$ levels that have not yet been studied. Elucidating such mechanisms is not only important for inkjet printing, but also essential for paper making, during which the $MC$ remains high. \\ \indent
\cite{paajanen2022nanoscale} showed that during single fiber \textit{hygro}-expansion up to a RH of 90\%, the hemi-cellulose inside the fibers softens between 40 and 60\% RH, whereas the cellulose micro-fibrils swell between 55 to 95\%. In \citep{vonk2021full, vonk2023frc} the transient dimensional changes accompanying these phenomena during \textit{hygro}-expansion (between 30 and 95\% RH) was characterized, revealing curves which are similar to conventional sorption isotherms \citep{fellers2007interaction, paajanen2022nanoscale}. However, characterization of paper above $MC$ levels associated with 95\% RH remains undiscovered, which according to \cite{salmen1987development} is the relevant $MC$ range during which phenomena occur that strongly affect the \textit{hygro}-expansivity of the paper sheet. \\ \indent
As generally known, restrained dried (RD) paper yields a significantly lower \textit{hygro}-expansion than freely dried (FD) paper \citep{salmen1987development, uesaka1992characterization, larsson2008influence, urstoger2020microstructure, vonk2023frc}. \cite{salmen1987development} stated that during the paper formation process when the paper is fully saturated, the so-called "dislocated regions", which make up \textsuperscript{$\sim$}50\% of the cellulose micro-fibrils, are soft and stretched through the restrained drying procedure, but subsequently hardened during drying from a "wet" to a 90\% RH state \citep{agarwal2013estimation}. For RD paper, during the restrained drying process, the cellulose molecules in the "dislocated regions" are prevented from contracting and therefore stretch and align. The aligned molecular configuration hardens, resulting in a largely reduced swelling of RD fibers when subjected to a $MC$ increase. In contrast, for FD handsheets, the cellulose molecules in the "dislocated regions" contract during drying, retaining its (more or less) amorphous configuration, resulting in a larger \textit{hygro}-expansion of FD fibers. The lower \textit{hygro}-expansivity of RD compared to FD fibers (between 30 and 90\% RH) was confirmed in \citep{vonk2023frc} and \citep{vonk2023res}, and it was indirectly shown (for hardwood fibers) that indeed the \textit{hygro}-expansion along the micro-fibrils was lower for RD fibers. Furthermore, the alignment of the "dislocated cellulose regions" due to an external force was confirmed using molecular dynamics simulations in \citep{khodayari2020tensile}. However, the dimensional change of FD and RD fibers and sheets inside the critical region in between 95\% RH and fully wet remains untested, as is the influence of the accompanying softening and alignment of the "dislocated cellulose regions" in the micro-fibrils \citep{salmen1987development, khodayari2020tensile}. \\ \indent
Therefore, the goal of this work is to investigate the transient \textit{hygro}- and \textit{hydro}-expansion of paper fibers and sheets, specifically in the region above 95\% RH and the fully wet state, and the effect of restrained drying. \citep{vonk2023res} serves as a basis, in which the \textit{hygro}-expansion of single fibers was tested before and after full water immersion, i.e. maximum $MC$. The purpose there was to test if RD fibers are able to "transform" into FD fibers upon water immersion, however, the transition from \textit{hygro}-expansion to full wetting and back was not studied in detail. The purpose of this work is to accurately measure the continuous expansion from the \textit{hygro}-expansion regime into the \textit{hydro}-expansion (full wetting) regime as well as the reversed transition, and do this for both handsheets and fibers extracted from these handsheets, for FD and RD, hardwood (HW) and softwood (SW). \\ \indent
Additionally, this work aims to predict the sheet scale \textit{hygro}- and \textit{hydro}-expansion using a dedicated bi-layer laminate model and asses the predictions compared to the experimental sheet expansions. Various advanced models have been proposed in the literature to predict the sheet \textit{hygro}-expansion from fiber characteristics, e.g. 2D fiber bond and homogenized network models \citep{bosco2015bridging, bosco2015predicting, bosco2017asymptotic, bosco2017hygro}, XFEM fiber models \citep{samantray2020level}, and 3D discrete network models \citep{motamedian2019simulating, brandberg2020role}. However, quasi-3D characterization of orthogonal inter-fiber bonds in \citep{vonk2023bonds} will show that a simple analytical bi-layer laminate model suffices to accurately predict the bending deformation of the bonded area, due to the \textit{hygro}-expansion differences of the two fibers. More insight can be gained when a set of measurements can be reproduced with a model containing only these essential ingredients. Therefore, here, the bi-layer laminate model will be extended to predict the sheet-scale \textit{hygro}-expansion. \\ \indent
The paper is structured as follows. First, the \textit{hygro}-expansion of fibers and sheets presented in \citep{vonk2023res} is briefly reviewed. Then, a novel sheet-scale \textit{hydro}-expansion measurement setup is presented which enables loading FD and RD, HW and SW handsheets from dry to fully saturated using liquid water, while tracking the associated dimensional changes. Next, a recently-developed, full-field fiber-scale \textit{hygro}-expansion method, based on global digital height correlation (GDHC), has been extended to track the transient evolution of the longitudinal, transverse, and shear strains of individual fibers up to the full wetting regime. This enables, for the first time, to monitor the complete time-dependent spectrum of the transition from \textit{hygro}- to \textit{hydro}-expansion of single fibers. With this new method, the transient \textit{hydro}-expansion of fibers extracted from the same handsheets as tested during the sheet-scale \textit{hydro}-expansion experiments is characterized. Collectively, enabling comparison of the \textit{hygro}- and \textit{hydro}-expansion of the fibers and sheets. Finally, an analytical model is derived and the \textit{hygro}- and \textit{hydro}-expansion predicted from the experimental fiber characteristics are compared to the sheet-scale experiments.

\section{Materials and Methods}
The sheet-scale \textit{hygro}-expansion and fiber-scale \textit{hygro}- and \textit{hydro}-expansion measurements are conducted using the recently-developed, advanced image correlation methods presented in \citep{vonk2023res}. Here, only the main aspects are therefore recalled. An experimental overview of all fibers and sheets of which the \textit{hygro}- and \textit{hydro}-expansion is characterized is given in Table \ref{tab:samples}.

\subsection{Handsheet preparation and hygro-expansion test method}
FD and RD, HW and SW handsheets of \textsuperscript{$\sim$}60 g/m\textsuperscript{2} were produced from bleached HW (Eucalyptus) and SW (mixture of Spruce and Pine) kraft pulp ($\kappa$ $<$2). The handsheets are cut to sheets of 6$\times$6 cm, and a random speckle pattern was applied to every handsheet using charcoal sticks, which is required for the Global Digital Image Correlation (GDIC) algorithm to enable identification of the in-plane \textit{hygro}-expansion. \\ \indent
The paper sheet-scale \textit{hygro}-expansion was subsequently monitored by applying the novel method proposed in \cite{vonk2023frc}, during which each handsheet was subjected to six linearly increasing RH cycles from 30$-$90$-$30\%, with a slope of 1\%/min. The collected images (obtained once every two minutes) are correlated using a GDIC framework with linear shape functions to obtain the evolution of the displacement field, which is converted to a strain measure (\textit{hygro}-expansion) \citep{neggers2016image}. Two FD and RD, HW and SW handsheets were characterized each.

\begin{figure}[t!]
	\centering
	\includegraphics[width=0.5\textwidth]{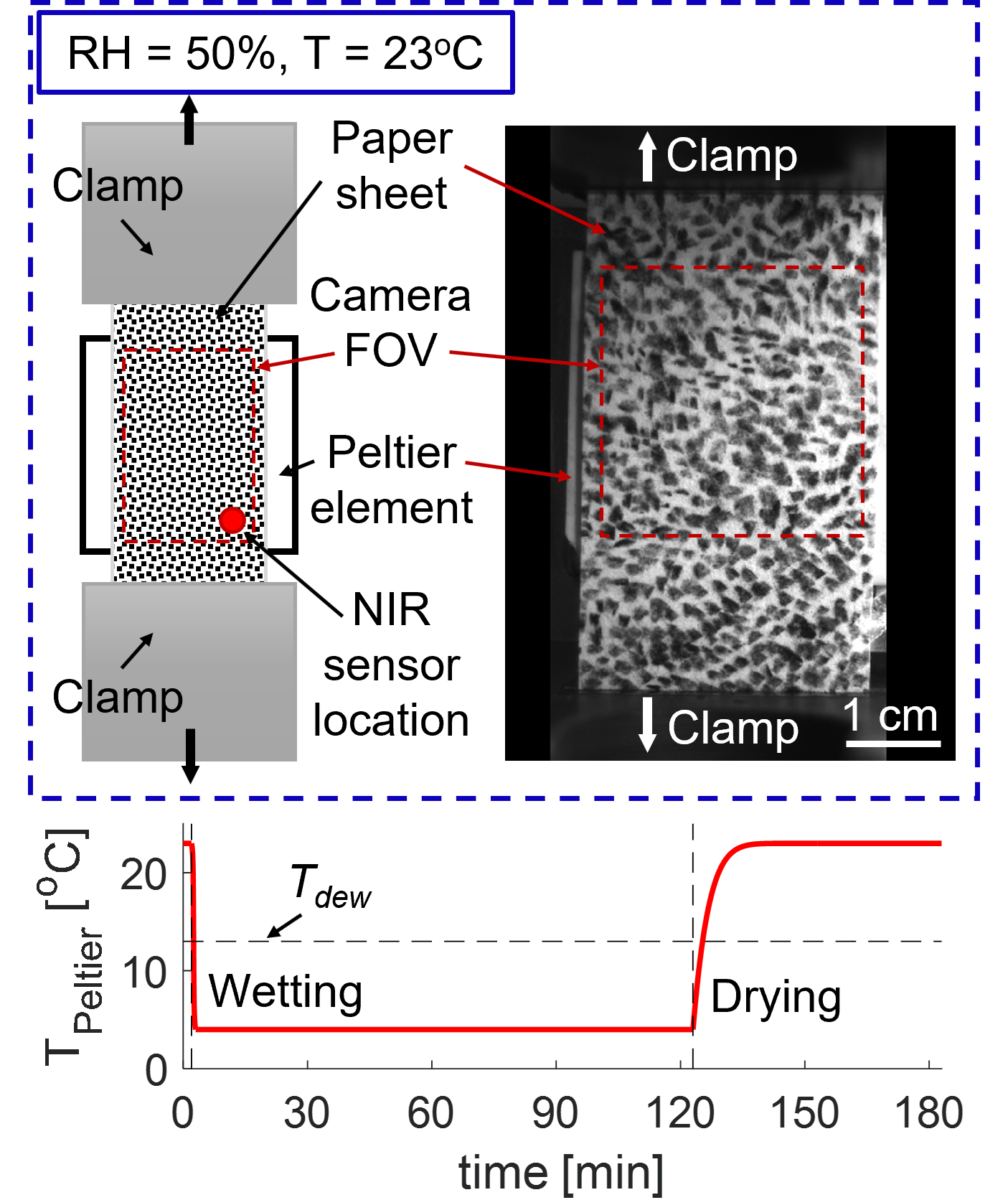}
	\caption{The novel paper sheet-scale \textit{hydro}-expansion setup, in which the paper sheet is clamped in a universal tensile tester to maintain a flat paper surface inside a 50\% and 23\textsuperscript{o}C climate. A Peltier element is located behind the paper sheet to cool down the handsheet, triggering condensation, consequently initiating \textit{hydro}-expansion. The patterned handsheet surface is captured by a camera, of which the images are correlated using a GDIC algorithm to obtain the transient deformation. $T_{peltier}$ is lowered to 4\textsuperscript{o}C, which is maintained for 120 minutes (wetting period) to attain full saturation. Afterwards, the Peltier element is shut off and $T_{peltier}$ rapidly increases to 23\textsuperscript{o}C (drying period), during which the handsheet is stabilized in 60 minutes. A near infrared sensor (NIR) is used to determine the average $MC$ of the handsheet at the specified position within the area of the Peltier element.}
	\label{fig:sheet_method}
\end{figure}
\subsection{A novel sheet-scale hydro-expansion method}
A novel setup, displayed in Figure \ref{fig:sheet_method}, is developed to characterize the \textit{hydro}-expansion of the same handsheets as used for the \textit{hygro}-expansion. The speckled handsheets are cut (halved) to strips of 6$\times$3 cm, and clamped by a universal tensile tester as explained below, which is located in a controlled climate of 50\% RH and 23\textsuperscript{o}C. To initiate the \textit{hydro}-expansion, a thermo-electric cooling (Peltier) element is located \textsuperscript{$\sim$}0.5$-$1 mm from the back of the paper sheet, as shown in Figure \ref{fig:sheet_method}, generating condensation when the Peltier element's temperature ($T_{Peltier}$) drops below the dew point temperature ($T_{dew}$), which is \textsuperscript{$\sim$}13\textsuperscript{o}C at 50\% RH. Therefore, to initiate the wetting cycle, $T_{Peltier}$ is lowered from 23 to 4\textsuperscript{o}C in \textsuperscript{$\sim$}15 minutes, as displayed in Figure \ref{fig:sheet_method}. The sheet first only exhibits \textit{hygro}-expansion, because insufficient condensation is formed to initiate \textit{hydro}-expansion. After \textsuperscript{$\sim$}10 minutes, condensation becomes large enough, i.e. \textsuperscript{$\sim$}1 mm, such that droplets become visible, yet the \textit{hydro}-expansion already initiated before this. $T_{Peltier}$ of 4\textsuperscript{o}C is chosen to generate the highest rate at which the water is introduced to the sheet, while not freezing the water due to possible temperature fluctuations. $T_{Peltier}$ is maintained at 4\textsuperscript{o}C for 120 minutes, which was sufficient to fully saturate the handsheet, as will be shown below. Afterwards, the Peltier element is shut off, through which $T_{Peltier}$ rapidly increases to 23\textsuperscript{o}C, hence initiating the drying period. Drying takes 60 minutes, which was sufficient to recover an equilibrium $MC$. \\ \indent
While $T_{Peltier}$ is altering, the handsheet surface is monitored every 10 seconds using a camera setup, and the transient handsheet \textit{hydro}-expansion is obtained by correlating the images using the same GDIC algorithm as used for the handsheet \textit{hygro}-expansion experiments. Note that due to the introduction of water, the brightness and contrast of the sheet surface strongly deteriorates. This was remedied by (i) adding an extra degree of freedom to the shape functions used in the GDIC, which allows for shifts in the gray value spectrum of the deformed image to match the undeformed image, and (ii) adopting a so-called incrementally 'updated' GDIC algorithm, in which each deformed image is correlated with its previous image instead of with the first image. Furthermore, throughout the experiment, the sheet is subjected to a constant force of 0.2 N, which is sufficient to straighten the paper sheet to minimize out-of-plane displacements, without inducing significant irreversible deformation. These out-of-plane displacements would otherwise induce artificial strains that are relatively large compared to the small strains expected during \textit{hydro}-expansion. Note that due to the applied force, the swelling becomes slightly anisotropic, which however is outweighed by the large strain errors that would result from out-of-plane deformations in case if no force was applied. To minimize the influence of the tensile strain, the strain in horizontal (free) direction is analyzed.\\ \indent
Finally, a near infrared (NIR) sensor is used to determine the average $MC$ of a discrete spot at the paper surface which lies within the Peltier element, as indicated in Figure \ref{fig:sheet_method}. To calibrate the NIR sensor, the tested sheets were fully saturated in water and subsequently dried on a scale (resolution: 10\textsuperscript{-4} gram) in a 50\% RH and 23\textsuperscript{o}C environment, while the NIR sensor output voltage and the handsheet weight are logged. Note that the NIR sensor is very sensitive to movement of the sample, hence care has been taken to find a NIR sensor position that yields the most stable $MC$ measurement. To this end, the NIR sensor is positioned 35 mm from the sheet, under and angle of 70\textsuperscript{o}, which is also used during the \textit{hydro}-expansion measurements. After drying, the dry weight of the paper was obtained by heating up the sheet to 105\textsuperscript{o}C. Combined, a linear trend was obtained between the voltage and $MC$, similar to \cite{larsson2009influence}.

\subsection{Preparation and hygro-expansion approach of the single fibers}
\begin{figure}[t!]
	\centering
	\includegraphics[width=0.5\textwidth]{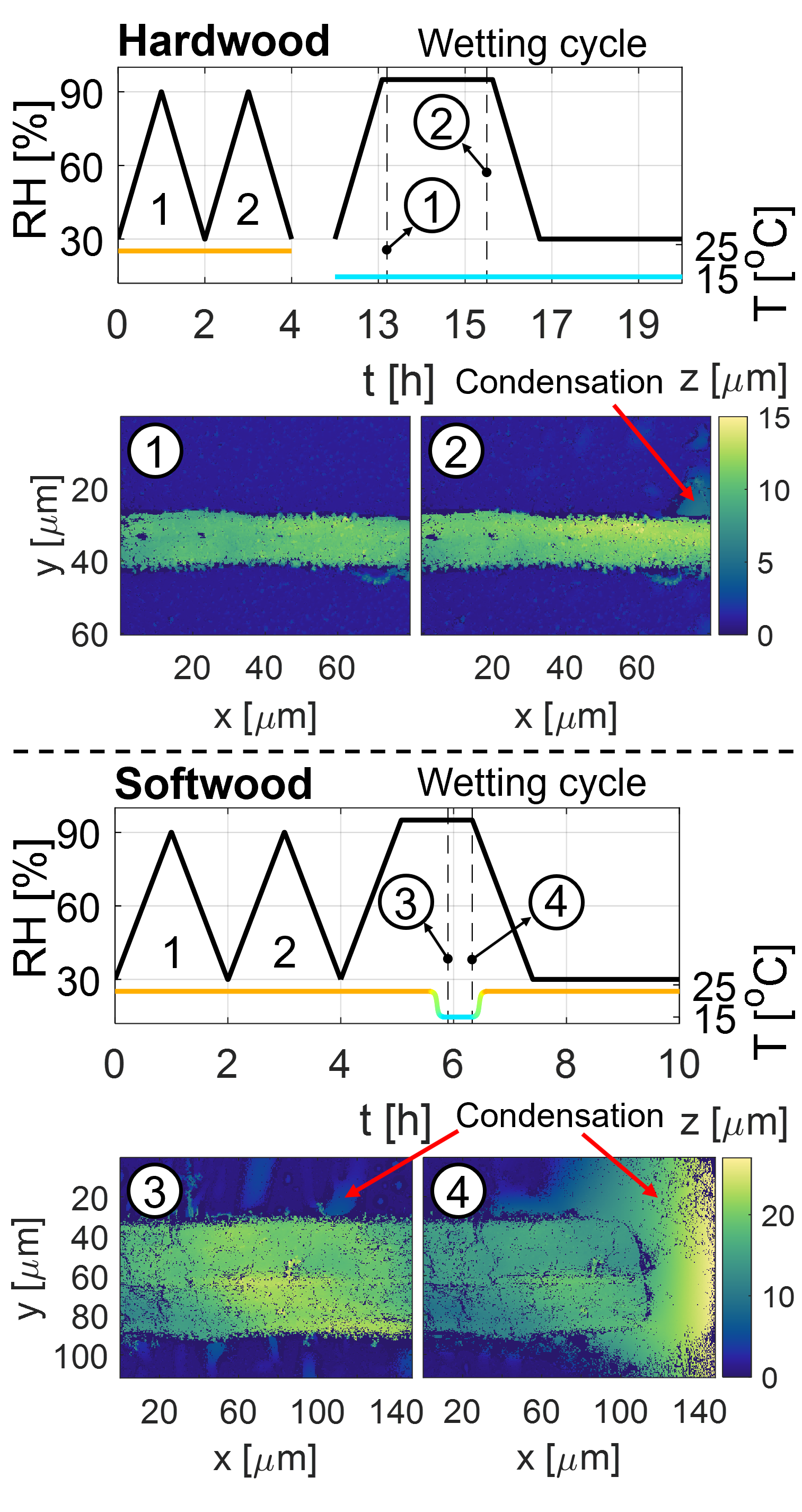}
	\caption{\textit{hygro}- and \textit{hydro}-expansion of fibers extracted from the FD and RD, HW and SW handsheets. After the \textit{hygro}-expansion cycles (cycles 1$-$2, 30$-$90$-$30\% RH), a wetting cycle is initiated during which the temperature of the fiber's substrate ($T$) and the RH are altered to generate condensation, as shown in topographies 2-4, consequently realizing fiber \textit{hydro}-expansion. The $RH$ and $T$ trajectories are different for HW and SW, i.e. for HW, $T$ is lowered at low RH after which the RH is raised to high level, resulting some droplets touching the fiber at the end of the wetting cycle (topography 2), whereas no droplet formation was visible when the high RH is reached (topography 1). For SW, $T$ is lowered while the RH was already at high level, resulting in an increased amount of droplets that touch the fibers (topography 3), however, these droplets grow out to a complete liquid front that runs over the fiber (topography 4), in contrast to the HW approach.}
	\label{fig:fiber_method}
\end{figure}

From the remainder of the handsheets five FD and nine RD HW fibers as well as ten FD and RD SW fibers were extracted. The fibers are prepared and tested following the method proposed by \cite{vonk2020robust}. First, the fibers are clamped onto a glass slide using two nylon wires to minimize rigid body translation, needed to keep the fiber in the microscope's field-of-view. Then, a 500 nm particle pattern is applied to the fiber using a dedicated mistification setup \citep{shafqat2021cool}, which is essential for the GHDC. \\ \indent
The fiber \textit{hygro}-expansion is characterized by changing the surrounding RH, while the fiber topography is captured with an optical profilometer using vertical scanning interferometry mode (\textit{Bruker NPFlex}) during swelling and shrinkage. The topographies are subsequently correlated using the GDHC algorithm to determine the longitudinal, transverse and shear surface strain fields during \textit{hygro}-expansion. \\ \indent
The prepared fiber specimens are mounted in a cooling stage, enabling a controlled temperature ($T$) close to the specimen, which is essential for realizing \textit{hydro}-expansion, as will be discussed below. Each fiber was subjected to two linearly increasing RH cycles from 30$-$90$-$30\% (cycle 1$-$2, \textit{hygro}-expansion) at 23\textsuperscript{o}C, as shown in Figure \ref{fig:fiber_method}, after which a wetting cycle initiates (\textit{hydro}-expansion). Note that all fibers were subjected to two 30$-$90$-$30\% RH (\textit{hygro}-expansion) cycles after the wetting cycle during the experiments discussed in \citep{vonk2023res}, which are outside the scope of this work and are not discussed here.

\subsection{Fiber hydro-expansion approach}
The full-field fiber \textit{hydro}-expansion is characterized within the same experiment as the full-field fiber \textit{hygro}-expansion, as displayed in Figure \ref{fig:fiber_method}. The fibers are subjected to a wetting cycle, during which the $MC$ is maximized. This is achieved by lowering the specimen temperature ($T$) to 15\textsuperscript{o}C and increasing the RH set-point to 95\%, and locally above \citep{fellers2007interaction}, triggering condensation ($T_{dew} \approx 22$\textsuperscript{o}C) at the fiber, realizing \textit{hydro}-expansion. The wetting cycle is, however, different for HW and SW, as visible in the trajectories given in Figure \ref{fig:fiber_method}. The reason behind this is that larger amount of condensation was required for the SW fibers (and possibly higher $MC$ levels) to understand the results presented in \citep{vonk2023res}, in which a more elaborate argumentation is given. \\ \indent
For the wetting cycle of the HW fibers, $T$ is decreased to 15\textsuperscript{o}C and stabilized at 30\% RH for 8 hours, before the wetting cycle initiates. The fiber surface is not monitored during this 8 hours, but, respectively, the last and first topography of the \textit{hygro}-expansion cycles and the wetting cycle are correlated to correct for possible strains. The RH is subsequently linearly increased from 30$-$95\% (slope of 1\%/min), and the 95\% RH set-point is kept constant for 3 hours. Visible condensation only starts to form at the end of the high RH period (topography 2), whereas no condensation was visible at the start (topography 1). During drying, the RH is linearly lowered to 30\% and is kept constant for 3 hours or longer until no condensation was visible anymore inside the climate chamber, and attain the same fiber state before and after the wetting cycle. \\ \indent
For the wetting cycle of the SW fibers, the RH is first linearly increased to 95\% and kept constant for 30 minutes, after which the temperature is lowered to 15\textsuperscript{o}C in approximately 15 minutes. This SW approach enabled the formation of an increased amount of (uncontrollable) condensation compared to the HW approach, which formed right after the change in $T$ (topography 3). The length of the wetting cycle varies per fiber, because $T$ and RH are simultaneously returned to their original value of, respectively, 23\textsuperscript{o}C and 30\% to counter the droplet formation when the condensation starts covering the fiber surface (topography 4), making the GDHC impossible. Again the experiment was stopped when no more condensation was visible. The SW approach generating significantly more and quicker condensation may be because the fiber resides in 95\% RH, which, upon cooling, directly condensates. Whereas for the HW approach, the condensation generation already occurred at lower RH levels and far from the fiber, because the humidifier's flow first passes the edge of cooling element (where the condensation starts to form) and feedback RH/$T$ sensor before reaching the fiber. In the following, for SW, the point at which $T$ is lowered up until the point where $T$ is increases again will be called the wetting period, and from there onward is called the drying period, as annotated in Figure \ref{fig:fiber_method}.

\subsection{Inter-fiber bond model}
\begin{figure}[t!]
	\centering
	\includegraphics[width=0.5\textwidth]{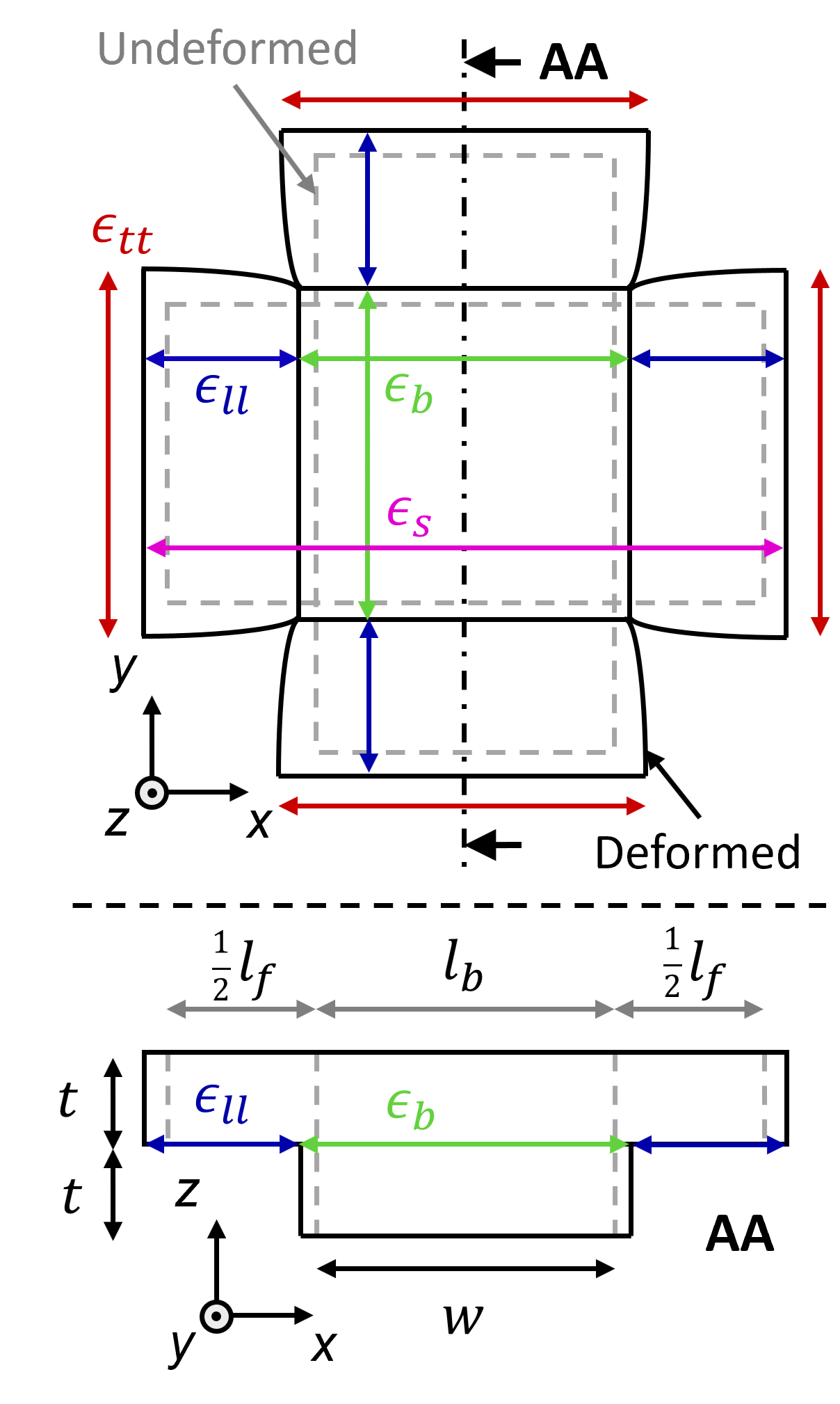}
	\caption{Schematic representation of the bi-layer laminate model used to predict the sheet-scale expansion ($\epsilon_s$) from the experimentally obtained fiber characteristics. The model consists of two orthogonal bonded fibers, in which the sheet expansion ($\epsilon_s$) is a combination of the longitudinal fiber strain ($\epsilon_{ll}$) in the freestanding arms, and the strain in the bonded area ($\epsilon_b$), which are weighted by the free fiber length ratio ($\eta$) given by $\l_f/(l_f+l_b)$.}
	\label{fig:bond_model}
\end{figure}

A schematic representation of the proposed inter-fiber bond model used to predict the sheet-scale \textit{hygro}- and \textit{hydro}-expansion from the experimentally obtained fiber \textit{hygro}- and \textit{hydro}-expansion characteristics is given in Figure \ref{fig:bond_model}. The model is based on a recently-developed inter-fiber model, which is elaborated and validated in a \citep{vonk2023bonds}, that is used to predict the strain curves through the thickness of the bonded area, and out-of-plane bending deformation during the \textit{hygro}-expansion of orthogonal inter-fiber bonds. Here, this model is adapted to predict sheet-scale expansion, with the main difference that the bending deformation is restricted to be zero, as bonds inside paper sheets are expected to be heavily constrained by neighboring fibers. \\ \indent
The sheet-scale expansion ($\epsilon_s$) is composed of the longitudinal fiber expansion in the freestanding arms ($\epsilon_{ll}$), and the bond strain ($\epsilon_b$), which results from the competition between the $\epsilon_{ll}$ of the first fiber and the transverse fiber expansion ($\epsilon_{tt}$) of the second fiber. $\epsilon_{ll}$ and $\epsilon_b$ are scaled via the free fiber length ratio ($\eta$), i.e.:
\begin{equation}
	\epsilon_s = \eta\epsilon_{ll}+(1-\eta)\epsilon_b.
\end{equation}
Here, $\eta$ equals $\l_f/(l_f+l_b)$, in which $l_f$ and $l_b$ are, respectively, the unbonded and bonded fiber length. To determine $\epsilon_b$, in \citep{vonk2023bonds} it will be shown that the bending deformation in the bonded area of isolated inter-fiber bonds is accurately described by a bi-layer laminate model based on classical laminate theory for thin plates \citep{pister1959elastic, reissner1961bending}. The rigorous derivation towards \textit{hygro}-expansion can be found in \citep{vonk2023bonds}. This model considers a linear variation of the in-plane strain components through the thickness coordinate of the bond ($z$), i.e. the strain profile in horizontal ($\epsilon_{xx}$) and vertical direction ($\epsilon_{yy}$), see Figure \ref{fig:bond_model} for the directions of $x$, $y$, and $z$, is given by:
\begin{equation}
	\label{eq:bend}
	\epsilon(z) = \epsilon_0+\kappa z,
\end{equation}
in which $\epsilon_0$ is the strain value at the interface between the two fibers, and $\kappa$ is the curvature. However, inside a paper sheet, the bending deformation is expected to be low due to the constraining neighboring fibers. Therefore, eliminating bending from the bi-layer model provides an estimate of the strain in the bond located inside a paper sheet. Then Equation \ref{eq:bend} simplifies to: 
\begin{equation}
	\label{eq:nobend}
	\epsilon(z) = \epsilon_0.
\end{equation}
Assuming two perpendicularly bonded fibers with equal characteristics, in terms of swelling ($\epsilon_{ll}$ and $\epsilon_{tt}$), geometry (thickness ($t$), width ($w$)), and mechanical properties ($E$, $\nu$), and free expansion ($F=M=0$), the inter-fiber bond strain ($\epsilon_b$) reduces to:
\begin{equation}
	\label{eq:nobendmodel}
	\begin{gathered}
		\epsilon_b = \frac{C_{11}\epsilon_{ll}+(C_{12}+C_{22})\epsilon_{tt}}{C_{11} + 	2C_{12}+C_{22}}, \\
		\text{with, } C_{11} = \frac{E_l}{1-\nu_{lt}\nu_{tl}}, \quad C_{12} = \frac{\nu_{tl}E_l}{1-\nu_{lt}\nu_{tl}},\text{ 	and} \quad C_{22} = \frac{E_t}{1-\nu_{lt}\nu_{tl}},
	\end{gathered}
\end{equation}
in which $E_l$ and $E_t$ are, respectively, the longitudinal and transverse stiffness, and $\nu_{lt}$ and $\nu_{tl}$ the Poisson's ratios. Note that due to the equal fiber dimensions, and the free expansion, Equation \ref{eq:nobendmodel} does not contain any geometrical parameters. Moreover, Equation \ref{eq:nobendmodel} reveals that $\epsilon_b$ does not depend on the absolute value of $E_l$ and $E_t$, but only the fiber stiffness ratio $E_t/E_l$. \\ \indent
Using Equation \ref{eq:nobendmodel}, the sheet-scale expansion can directly be predicted, upon inserting the experimentally obtained fiber swelling properties, i.e. $\epsilon_{ll}$ and $\epsilon_{tt}$, assuming a free fiber length of \textsuperscript{$\sim$}50\% \citep{wernersson2014characterisations, borodulina2016extracting, urstoger2020microstructure}, and adopting the fiber stiffness characteristics from literature with $\nu_{lt}=\nu_{tl}$ equal to 0.022 \citep{magnusson2013numerical, brandberg2020role, czibula2021transverse}. $E_t/E_l$ is different for FD and RD, HW and SW fibers \citep{jentzen1964effect}, therefore a range of values for $E_t/E_l$ between 6 and 11 is studied here, covering the relevant range of values found in literature \citep{magnusson2013numerical, brandberg2020role, czibula2021transverse}. 

\section{Results and discussion}
Many different strain quantities are compared in this paper, therefore the following notation is consistently adopted in this work: $\epsilon^{\gamma/\delta+/-}_{s/ll/tt/lt}$, in which superscript \textsuperscript{$\gamma$} versus \textsuperscript{$\delta$} denote \textit{hygro}- versus \textit{hydro}-expansion (referring to the Greek $\upsilon\bm{\gamma}\rho o\sigma\kappa o\pi\iota\kappa \acute o\varsigma$ (\textit{hygro}scopic) and $\upsilon\bm{\delta}\rho o\sigma\kappa o\pi\iota\kappa \acute o\varsigma$ (\textit{hydro}scopic)); superscript \textsuperscript{+} versus \textsuperscript{-} denote, respectively, wetting versus drying; and subscript $_s$ denotes sheet-scale values, whereas the fiber-scale strain components are denoted with subscripts $_{ll}$, $_{tt}$, and $_{lt}$, corresponding to, respectively, longitudinal, transverse and shear strain. All of the above strains notations are variables, whereas the absolute strain changes during wetting and drying, which are constant values, are denoted using a $\Delta$ under the strain symbol, i.e. $\underset{^\Delta}{\epsilon}$. Finally, a bar over the strain symbol, i.e. $\bar{\epsilon}$, denotes the average value of a measurement ensemble. 

\subsection{Fiber and sheet hygro-expansion}
\begin{figure}[t!]
	\centering
	\includegraphics[width=0.5\textwidth]{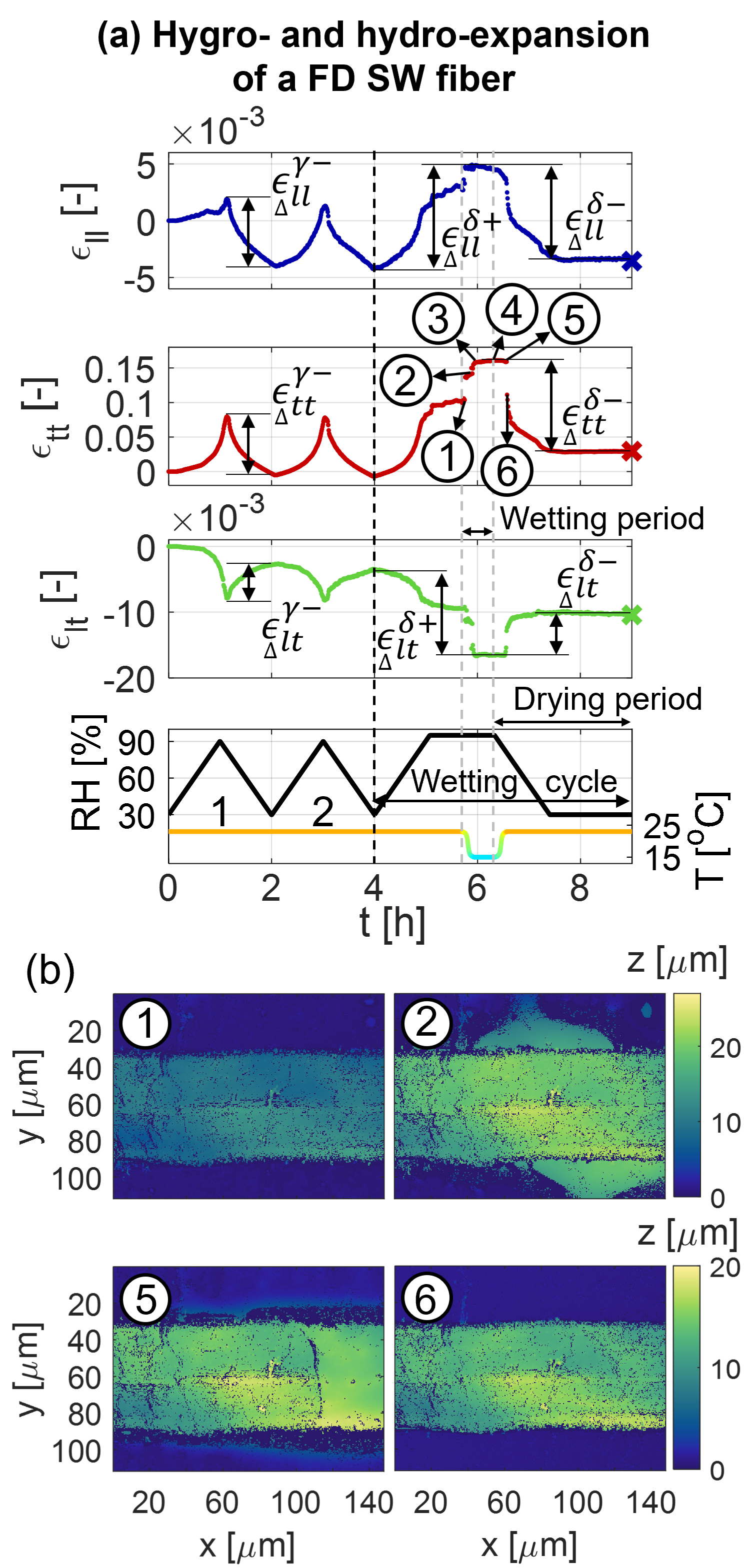}
	\caption{(a) The transient longitudinal ($\epsilon_{ll}$), transverse ($\epsilon_{tt}$), and shear strain ($\epsilon_{lt}$) during \textit{hygro}- (cycles 1$-$2) and \textit{hydro}-expansion (wetting cycle) of a FD SW fiber. The strains change during drying from 90 to 30\% (\textit{hygro}-expansion, cycles 1$-$2), annotated as $\underset{^\Delta}{\epsilon}$$^{\gamma-}_{ll}$, $\underset{^\Delta}{\epsilon}$$^{\gamma-}_{tt}$, and $\underset{^\Delta}{\epsilon}$$^{\gamma-}_{lt}$ are extracted. (b) Four topographies, as annotated in the $\epsilon_{tt}$ curve (a), are used to display the phenomena during wetting of the SW fibers, in order to better understand the \textit{hydro}-expansion response. Note that topography 3 and 4 are given in Figure \ref{fig:fiber_method}. The strain changes during the wetting cycle for wetting and drying, respectively, $\underset{^\Delta}{\epsilon}$$^{\delta+}_{ll}$, $\underset{^\Delta}{\epsilon}$$^{\delta+}_{tt}$, and $\underset{^\Delta}{\epsilon}$$^{\delta+}_{lt}$, and $\underset{^\Delta}{\epsilon}$$^{\delta-}_{ll}$, $\underset{^\Delta}{\epsilon}$$^{\delta-}_{tt}$, and $\underset{^\Delta}{\epsilon}$$^{\delta-}_{lt}$, are also extracted. The crosses (at $t=9$ hours) in (a) indicate the fiber strain after equilibrating the RH and $T$ to, 30\% and 23\textsuperscript{o}C respectively, equal to the start of the experiment (at $t=0$ hours).}
	\label{fig:fiber_trend}
\end{figure}

\begin{figure}[t!]
	\centering
	\includegraphics[width=0.5\textwidth]{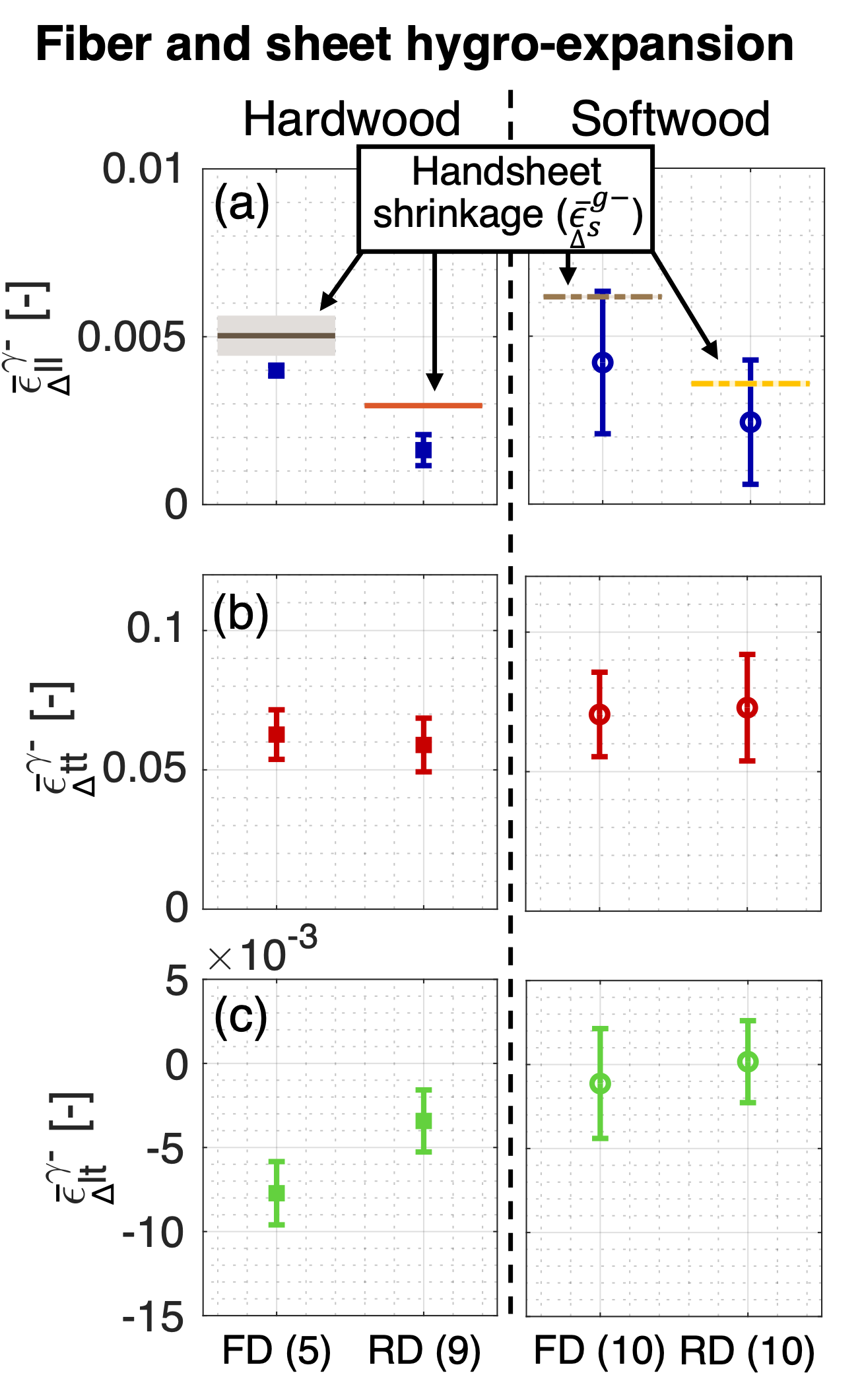}
	\caption{The average (a) longitudinal and (b) transverse shrinkage, and (c) shear strain change during drying from 90 to 30\% RH of cycles 1$-$2 of all fibers, i.e. $\underset{^\Delta}{\bar{\epsilon}}$$^{\gamma-}_{ll}$, $\underset{^\Delta}{\bar{\epsilon}}$$^{\gamma-}_{tt}$, and $\underset{^\Delta}{\bar{\epsilon}}$$^{\gamma-}_{lt}$, including their standard deviation. The average handsheet shrinkage ($\underset{^\Delta}{\bar{\epsilon}}$$^{\gamma-}_{s}$ given in Table \ref{tab:exp_sheet_hygro}) is added to the $\underset{^\Delta}{\bar{\epsilon}}$$^{\gamma-}_{ll}$ plot to compare.}
	\label{fig:fiber_sheet_hygro}
\end{figure}

Figure \ref{fig:fiber_trend} (a) shows a typical fiber-scale measurement of the longitudinal, transverse and shear strain of a FD SW fiber, monitored during 2 \textit{hygro}-expansion cycles and 1 \textit{hydro}-expansion cycle (with corresponding topographies in (b)). For now, only the \textit{hygro}-expansion in Figure \ref{fig:fiber_trend} is discussed. The fibers' longitudinal and transverse shrinkage, and shear strain change during drying from 90 to 30\% RH (\textit{hygro}-expansion) for cycles 1$-$2, i.e. $\underset{^\Delta}{\epsilon}$$^{\gamma-}_{ll}$, $\underset{^\Delta}{\epsilon}$$^{\gamma-}_{tt}$, and $\underset{^\Delta}{\epsilon}$$^{\gamma-}_{lt}$, as annotated in Figure \ref{fig:fiber_trend}, are extracted for each fiber. The average shrinkage considering cycles 1$-$2 of all fibers, i.e. $\underset{^\Delta}{\bar{\epsilon}}$$^{\gamma-}_{ll}$, $\underset{^\Delta}{\bar{\epsilon}}$$^{\gamma-}_{tt}$, and $\underset{^\Delta}{\bar{\epsilon}}$$^{\gamma-}_{lt}$, including their standard deviation, are displayed for FD and RD, HW and SW in Figure \ref{fig:fiber_sheet_hygro}, and given in Table \ref{tab:fiber_hygro}. \\ \indent
From the transient sheet-scale \textit{hygro}-expansion curve of two FD and RD, HW and SW handsheets, the shrinkage for each drying slope from 90 to 30\% RH is extracted. The average shrinkage considering both sheets per handsheet type ($\underset{^\Delta}{\bar{\epsilon}}$$^{\gamma-}_{s}$), including its standard deviation were plotted in the top row of Figure \ref{fig:fiber_sheet_hygro}, and given in Table \ref{tab:exp_sheet_hygro}. \\ \indent
Regarding the handsheet \textit{hygro}-expansion, Figure \ref{fig:fiber_sheet_hygro} shows that $\underset{^\Delta}{\bar{\epsilon}}$$^{\gamma-}_{s}$ is significantly larger for the FD compared to RD, similar to literature \citep{uesaka1992characterization, nanko1995mechanisms, larsson2008influence, urstoger2020microstructure, vonk2023frc}, and $\underset{^\Delta}{\bar{\epsilon}}$$^{\gamma-}_{s}$ of the SW handsheets is slightly larger than for their HW counterparts. Regarding the fiber \textit{hygro}-expansion, $\underset{^\Delta}{\bar{\epsilon}}$$^{\gamma-}_{ll}$ is also significantly larger for FD compared to RD, similar to the corresponding handsheets. This, combined with the fact that the longitudinal fiber swelling strongly affects the dimensional stability of paper \citep{uesaka1994general, brandberg2020role, motamedian2019simulating}, indicates that the sheet-scale \textit{hygro}-expansion differences between FD and RD are predominantly caused by the differences at the fiber level, in contrast to reports in the literature attributing the FD/RD differences to geometrical differences in the inter-fiber bonds \citep{uesaka1994general, larsson2008influence}. The importance of the longitudinal fiber \textit{hygro}-expansion is further supported by the observation that, for all handsheets, $\underset{^\Delta}{\bar{\epsilon}}$$^{\gamma-}_{s}$ is only slightly larger than $\underset{^\Delta}{\bar{\epsilon}}$$^{\gamma-}_{ll}$, indicating that the transverse fiber \textit{hygro}-expansion contribution to the sheet scale is small, as observed in the \textit{hygro}-expansion experiments on isolated inter-fiber bonds in \citep{vonk2023bonds}, and models of the network \textit{hygro}-expansion \citep{motamedian2019simulating, brandberg2020role}.

\subsection{Sheet-scale hydro-expansion}
\begin{figure}[t!]
	\centering
	\includegraphics[width=\textwidth]{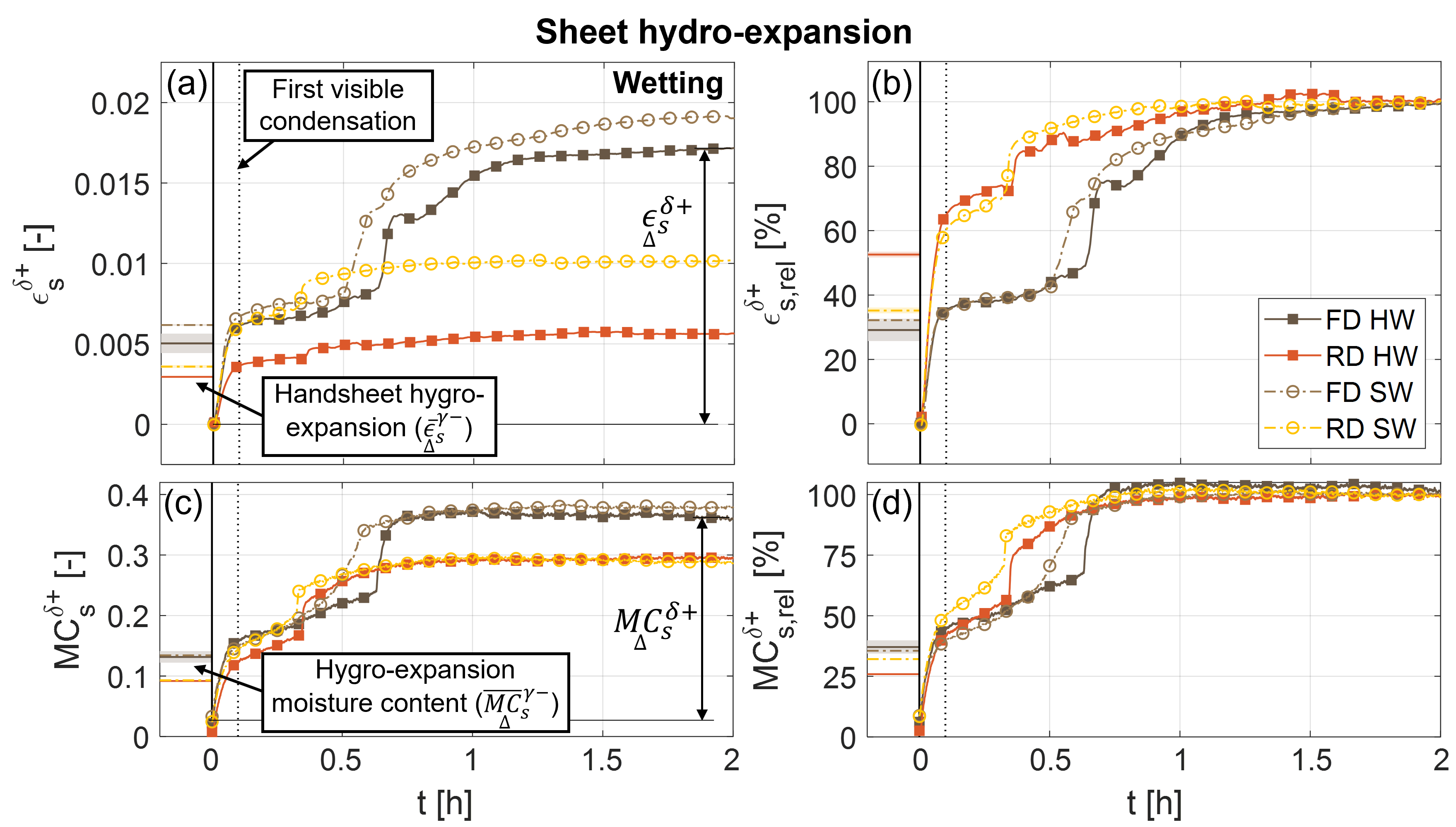}
	\caption{The transient sheet-scale \textit{hydro}-expansion and $MC$ during wetting of the sheet-scale \textit{hydro}-expansion experiment, i.e., respectively, $\epsilon^{\delta+}_s$ and $MC^{\delta+}_s$. $\epsilon^{\delta+}_s$ and $MC^{\delta+}_s$ are divided by, respectively, their maximum value at the end of the wetting cycle (at $t = 2$ hours), to find the $\epsilon^{\delta+}_{s,rel}$ and $MC^{\delta+}_{s,rel}$ curves, enabling a better comparison between the handsheets. The handsheet shrinkage during the \textit{hygro}-expansion experiments ($\underset{^\Delta}{\bar{\epsilon}}$$^{\gamma-}_{s}$ in Figure \ref{fig:fiber_sheet_hygro}) and $\underset{^\Delta}{\bar{MC}}$$^{\gamma-}_{s}$ are added to highlight the magnitude differences between the \textit{hygro}- and \textit{hydro}-expansion experiments. The first condensation was visible after 10 minutes for all handsheets, marked by the dotted line in the figures. The total changes in $\epsilon^{\delta+}_s$ and $MC^{\delta+}_s$, i.e., respectively, $\underset{^\Delta}{\epsilon}$$^{\delta+}_{s}$ and $\underset{^\Delta}{MC}$$^{\delta+}_{s}$, are extracted.}
	\label{fig:sheet_hydro}
\end{figure}

\begin{figure}[t!]
	\centering
	\includegraphics[width=0.5\textwidth]{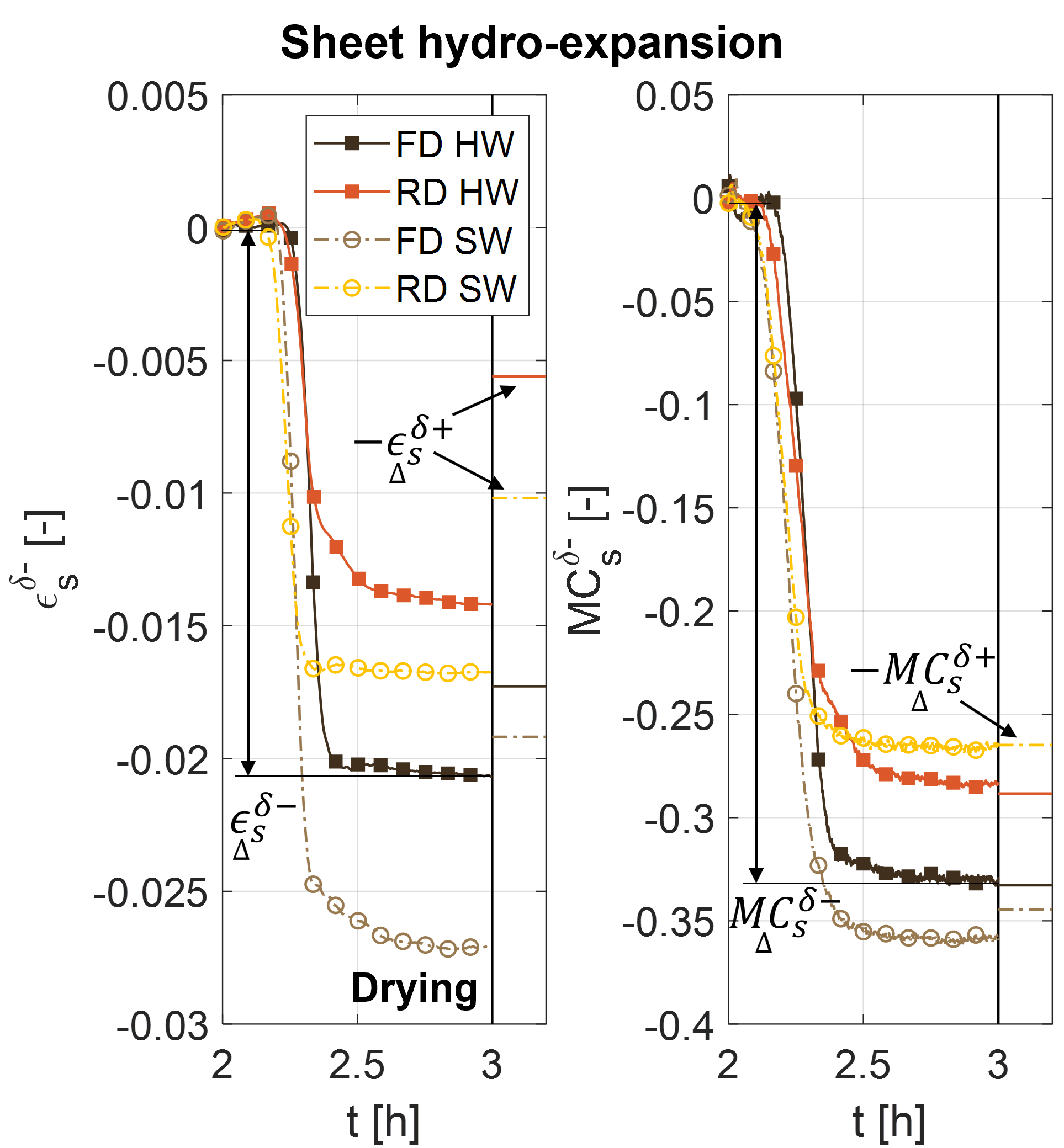}
	\caption{The transient sheet-scale \textit{hydro}-expansion and $MC$ during drying of the sheet-scale \textit{hydro}-expansion experiment, i.e., respectively, $\epsilon^{\delta-}_s$ and $MC^{\delta-}_s$. $\epsilon^{\delta-}_s$ and $MC^{\delta-}_s$ start at 0 to display the shrinkage of the sheet from wet to dry, to assist in the visual comparison of the curves. The negative values of $\underset{^\Delta}{\epsilon}$$^{\delta+}_{s}$ and $\underset{^\Delta}{MC}$$^{\delta+}_{s}$ are added right of the curves to reveal the release of irreversible dried-in strains, showing that $MC^\delta_s$ at the start and end of the experiment are almost the same ($\underset{^\Delta}{MC}$$^{\delta-}_{s}$$=$$\underset{^\Delta}{MC}$$^{\delta+}_{s}$), which is clearly not the case for $\epsilon^{\delta}_s$ ($\underset{^\Delta}{\epsilon}$$^{\delta-}_{s}$$\neq$$\underset{^\Delta}{\epsilon}$$^{\delta+}_{s}$). The final change in $\epsilon^{\delta-}_s$, i.e. $\underset{^\Delta}{\epsilon}$$^{\delta-}_{s}$, is extracted.}
	\label{fig:sheet_hydro_drying}
\end{figure}

The sheet-scale \textit{hydro}-expansion and $MC$ evolution of a FD and RD, HW and SW handsheet during the wetting cycle of the sheet-scale \textit{hydro}-expansion experiments, i.e., respectively, $\epsilon^{\delta+}_s$ and $MC^{\delta+}_s$, are displayed in Figure \ref{fig:sheet_hydro} (a, c). To enable a better comparison between the handsheets, $\epsilon^{\delta+}_s$ and $MC^{\delta+}_s$ are also plotted relative to the maximum value found at the end of the wetting period (i.e. at $t=2$ hours), i.e. $\epsilon^{\delta+}_{s,rel}$ and $MC^{\delta+}_{s,rel}$ in Figure \ref{fig:sheet_hydro} (b, d). The average sheet-scale shrinkage during \textit{hygro}-expansion ($\underset{^\Delta}{\bar{\epsilon}}$$^{\gamma-}_{s}$), given in Figure \ref{fig:fiber_sheet_hygro} (a), has been added to the left of the curves for comparison. Additionally, an approximation of the $\underset{^\Delta}{\bar{MC}}$$^{\gamma-}_{s}$ is obtained by looking up the $MC^{\delta+}_s$ value corresponding to the $\underset{^\Delta}{\bar{\epsilon}}$$^{\gamma-}_{s}$ value in the $\epsilon^{\delta+}_s$-$MC^{\delta+}_s$ curve, which are added to the $MC^{\delta+}_s$ plots. Note that this is an approximated $\underset{^\Delta}{\bar{MC}}$$^{\gamma-}_{s}$, because $MC^\gamma_s$ was not monitored during the \textit{hygro}-expansion experiments, only during the \textit{hydro}-expansion experiments. \\ \indent
First of all, the novel sheet-scale \textit{hydro}-expansion method is sufficiently robust to characterize the evolution of the sheet expansion from dry to fully wet with high precision, especially considering the large brightness and contrast deterioration of the images during the experiments. The wetting was achieved relatively uniformly over the paper surface, because no localized hot spots were visible in the converged correlation residual field from the GDHC algorithm. Moreover, $\epsilon^{\delta+}_s$ and $MC^{\delta+}_s$ display similar curves, indicating that both measurements were performed consistently, because the moisture-induced dimensional change of FD paper should, approximately, depend linearly on the $MC$ \citep{uesaka1992characterization, larsson2008influence}. Furthermore, the handsheets reach an equilibrium much later for wetting than drying due to slower mass transfer during wetting, similar to the sheet and fibers \textit{hygro}-expansion measurements in the literature \citep{niskanen1997dynamic, vonk2021full}. Finally, note that the sheets first exhibit \textit{hygro}-expansion before the condensation droplets become visible after approximately 10 minutes, during which \textit{hydro}-expansion sets in. After this point, most handsheets swell a factor 2 more compared to the final swelling after \textit{hygro}-expansion from 30 to 90\% (Figure \ref{fig:sheet_hydro}), whereby both $\epsilon^{\delta+}_s$ and $MC^{\delta+}_s$ converge (approximately) to a plateau value. This indicates that the sheet \textit{hydro}-expansion was successfully characterized in between the 90\% RH and the fully saturated regime.\\ \indent
Similar to the sheet \textit{hygro}-expansion experiments (Figure \ref{fig:fiber_sheet_hygro}), in the \textit{hydro}-expansion experiments, the FD sheets exhibit a larger $\epsilon^{\delta+}_s$ at the end of the wetting cycle than the RD handsheets, and the SW handsheets swell more than their corresponding HW counterparts. The ultimate swelling resulting from wetting of the \textit{hydro}-expansion cycle, i.e., $\underset{^\Delta}{\epsilon}$$^{\delta+}_{s}$ annotated in Figure \ref{fig:sheet_hydro}, is extracted and given in Table \ref{tab:exp_sheet_hydro}. Interestingly, the transient $\epsilon^{\delta+}_s$ and $MC^{\delta+}_s$ display significant differences from conventional \textit{hygro}-expansion curves, i.e., all handsheets show an abrupt transition (step) in $\epsilon^{\delta+}_s$ and $MC^{\delta+}_s$ around $t=0.5$ hours. This step is significantly larger and occurs later for FD compared to the RD handsheets. This step becomes even better visible for the $\epsilon^{\delta+}_{s,rel}$ and $MC^{\delta+}_{s,rel}$ curves. At first glance, it may seem that this step characterizes the transition from \textit{hygro}- to \textit{hydro}-expansion, during which liquid water directly ingresses in the sheet and hence the sheet swells quicker. However, the first water droplets were already visible after 10 minutes, indicated by the black, dotted line in Figure \ref{fig:sheet_hydro}), and the step occurs much later. Moreover, the FD and RD sheets reveal distinctly different timescales for this step, whereas SW and HW do not show a clear difference, which cannot be trivially explained if the time scale is governed by the absorption of water in the network. Additionally, it is known from literature that the gross network water absorption should occur much faster, i.e. in the order of seconds/minutes \citep{niskanen1997dynamic, jajcinovic2018influence}, which corresponds with the initial increase in $MC^{\delta+}_s$ that occurs within the first 10 minutes, where the slightly longer timescale is caused by the non-instantaneous temperature drop. Note also that this initial increase occurs with precisely the same timescale for FD and RD and for SW and HW, see Figure \ref{fig:sheet_hydro} (d), as is to be expected for the uptake of water in the network. The fact that the type of drying process rather than the type of (SW or HW) fibers mainly affects the timescale of the step (which occurs after the initial increase) indicates that the drying process can be a controlling mechanism. Based on this insight, a hypothesis for this process is proposed below, for which the sheet \textit{hydro}-expansion during drying and the fiber \textit{hydro}-expansion need to be rationalized first. \\ \indent
The evolution of the sheet-scale \textit{hydro}-expansion and $MC$ during drying of the sheet-scale \textit{hydro}-expansion experiment, i.e., respectively, $\epsilon^{\delta-}_s$ and $MC^{\delta-}_s$, combined with the negative values of $\underset{^\Delta}{\epsilon}$$^{\delta+}_{s}$ and $\underset{^\Delta}{MC}$$^{\delta+}_{s}$ are given in Figure \ref{fig:sheet_hydro_drying}. The final shrinkage after drying, i.e. $\underset{^\Delta}{\epsilon}$$^{\delta-}_{s}$, is extracted and given in Table \ref{tab:exp_sheet_hydro}. Similar to the wetting cycles, the RD and HW handsheets shrink less than, respectively, the FD and SW handsheets. Furthermore, all sheets show roughly the same $MC^\delta_s$ before and after the \textit{hydro}-expansion cycle, indicating that the experiment was conducted properly, such that the wetting and drying strains can be quantitatively compared. Similar to the \textit{hygro}-expansion experiments, the RD handsheets exhibit a larger release of irreversible dried-in strain relative to their swelling cycle ($\underset{^\Delta}{\epsilon}$$^{\delta-}_{s}$$/$$\underset{^\Delta}{\epsilon}$$^{\delta+}_{s}$), i.e. for the RD HW and SW sheets, the shrinkage magnitude is, respectively, 2.57 and 1.64 times larger than $\underset{^\Delta}{\epsilon}$$^{\delta+}_{s}$, whereas for the FD HW and SW handsheets, this is only 1.21 and 1.41. \\ \indent 
In \citep{vonk2023res} it was found that the FD fibers exhibit a significantly larger \textit{hygro}-expansivity than the RD fibers, and the RD HW fibers can "transform" to HW fibers exhibiting the \textit{hygro}-expansion characteristics of FD HW fiber, while the RD SW fibers cannot. Considering the sheet-scale shrinkage magnitudes here, the difference between the RD and FD handsheets is much larger for HW than SW, suggesting that the fiber level "transformation" for RD to FD HW fibers also occurs for RD HW sheets. However, note that this sheet-scale "transformation" is not fully effective because the shrinkage of the FD HW handsheet remains larger than the RD HW handsheet in Figure \ref{fig:sheet_hydro_drying}. To test if the "transformation" really occurs at the sheet level, the hygro-expansion of the handsheets tested during the \textit{hydro}-expansion experiments should be characterized afterwards. Furthermore, as generally known, the elastic stiffness of RD handsheets is larger than FD handsheets \citep{urstoger2020microstructure, alzweighi2021influence}. Hence, to test if not only the hygroscopic properties of RD handsheets "transform" after wetting, the elastic stiffness of the FD and RD handsheets before and after the the \textit{hydro}-expansion cycle should be obtained. 

\subsection{Fiber hydro-expansion}
\begin{figure}[b!]
	\centering
	\includegraphics[width=0.5\textwidth]{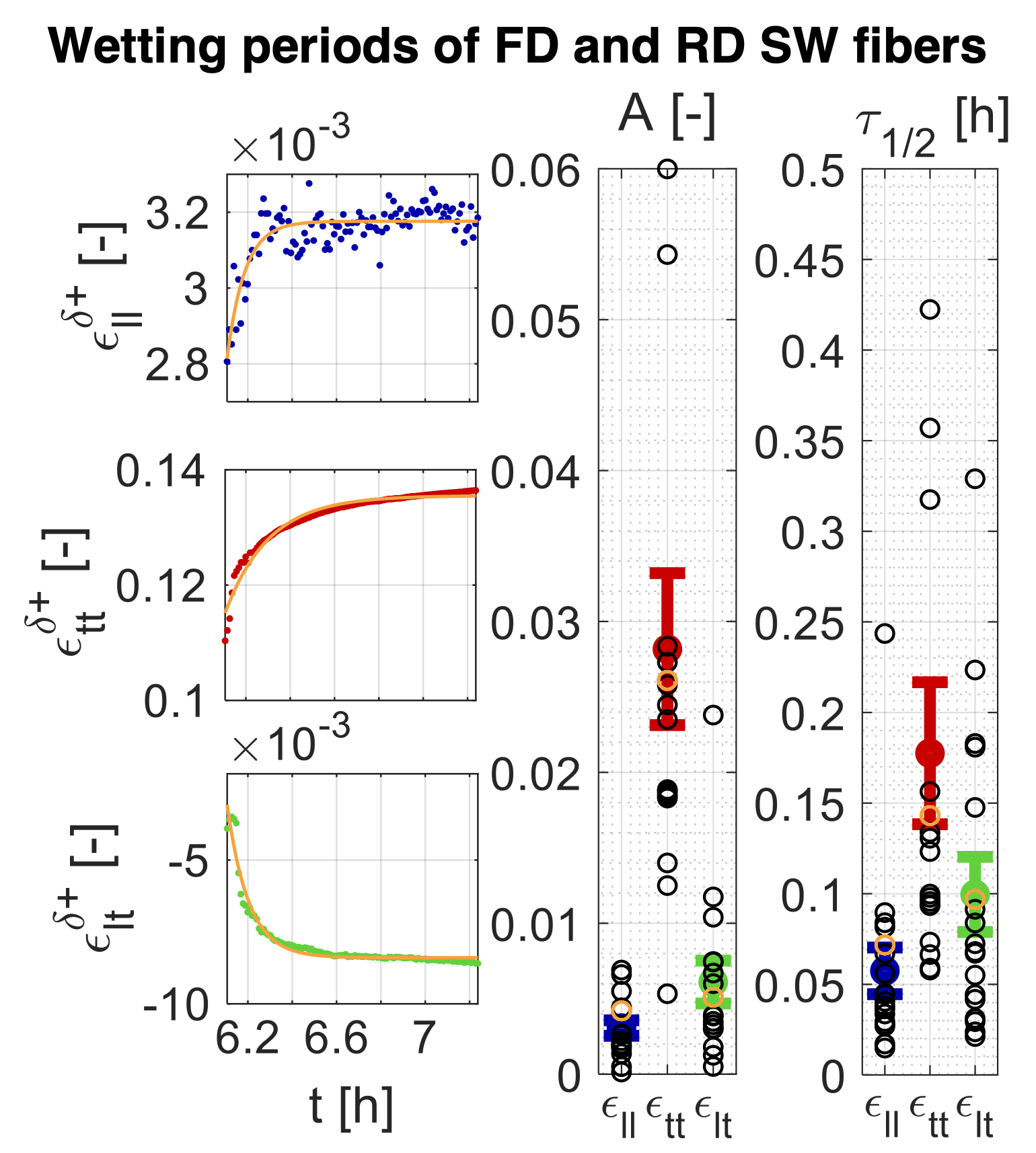}
	\caption{Strain step amplitude ($A$) and saturation half-times ($\tau_{1/2}$) of the longitudinal, transverse and shear strain curve of all SW fibers during the wetting period (annotated in the wetting cycle in Figure \ref{fig:fiber_trend}) after the hypothesized softening of the "dislocated regions" in the cellulose micro-fibrils. $A$ and $\tau_{1/2}$ are obtained by fitting the strain curves using $Ae^{1-\frac{t}{\tau_{1/2}}}+B$, where the constants $A$ and $B$ are given in the $\epsilon_{ll}$ plot. In the ensemble plots, the three values plotted in orange open circles correspond to the fits in the three separate strain plots shown on the left, and the black symbols open circles represent the values of the other fibers. The mean values of $A$ and $\tau_{1/2}$ are shown by the solid circles, plotted together with an error bar denoting the standard deviation of the mean.}
	\label{fig:timescales}
\end{figure}

Let us remember the combined transient \textit{hygro}- and \textit{hydro}-expansion response of a FD SW fiber given in Figure \ref{fig:fiber_trend} (a). First of all, the fiber swells significantly more during the wetting cycle compared to RH cycles 1$-$2, indicating that higher $MC$ levels are attained. Next, the \textit{hydro}-expansion is initiated just after the $T$ is lowered, inducing the tiny droplets observed in topography 1 in Figure \ref{fig:fiber_trend} (b). Interestingly, topography 2 displays a water droplet residing under the fiber, which disappeared in the next topography and did not reappear in the following topographies (topography 3 in Figure \ref{fig:fiber_method}), indicating that the fiber was able to fully absorb the (relatively large) droplet. Furthermore, for all SW fibers, the fiber shrinkage is significantly delayed from the RH curve because condensation is present around the SW fibers at the start of the drying period, requiring time to evaporate. During the drying period, the FD SW fiber shows a large jump in $\epsilon^{\delta-}_{tt}$, which is less prominently visible for $\epsilon^{\delta-}_{ll}$ and $\epsilon^{\delta-}_{lt}$. All other FD and RD SW fibers displayed similar steps. The formation of a water layer in between the fiber and glass (topography 5) most likely fixates the fiber to the substrate due to capillary forces, similar to \cite{page1963transverse}. This layer disappeared just after the strain jump (topography 6 in Figure \ref{fig:fiber_trend}). The HW fibers did not show this jump because no water layer was formed in these tests, resulting in introducing less condensation to the HW fibers compared to the SW fibers. \\ \indent
\begin{figure}[b!]
	\centering
	\includegraphics[width=1\textwidth]{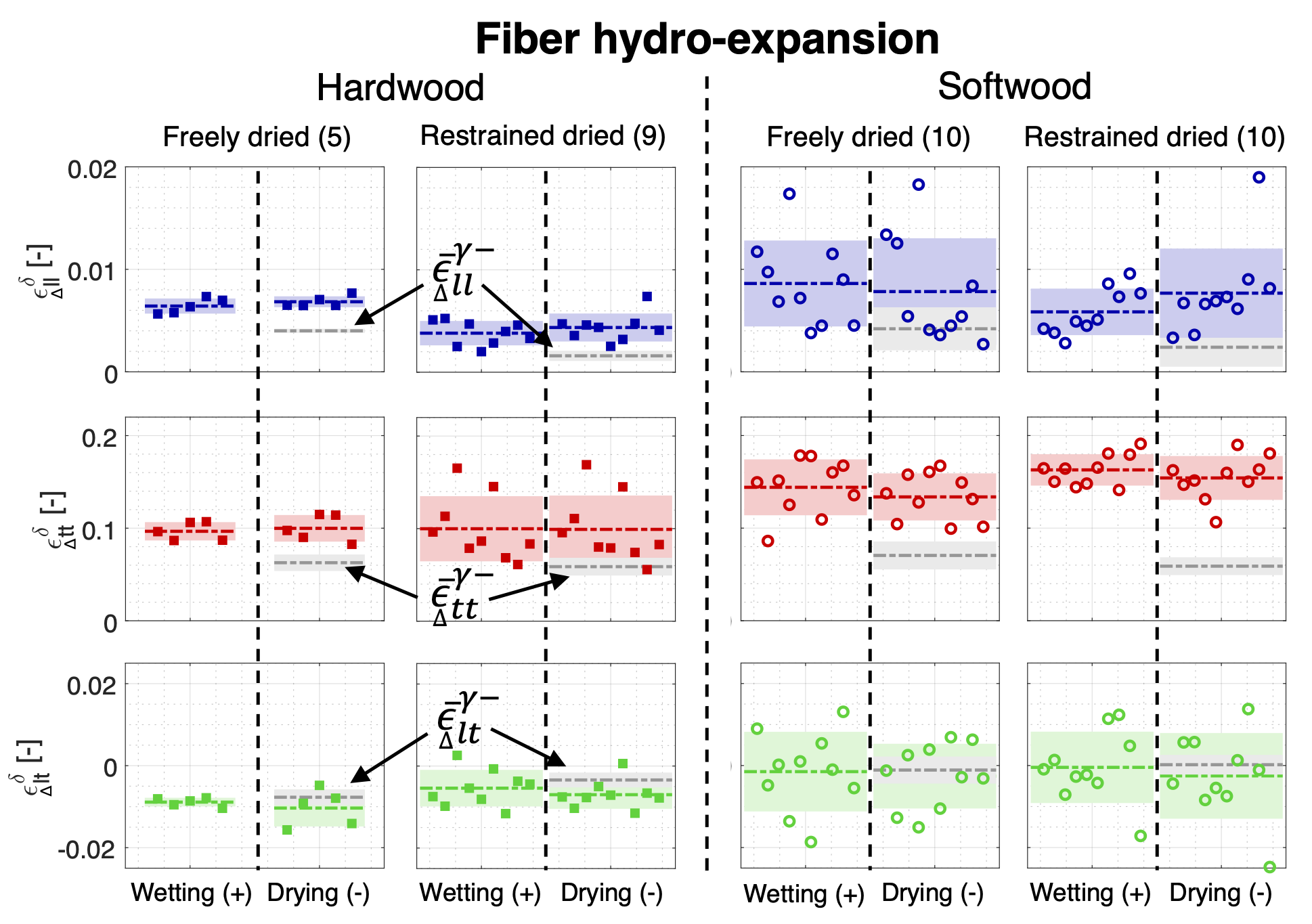}
	\caption{The maximum longitudinal, transverse and shear strain changes per fiber for wetting and drying during the wetting cycle annotated as, respectively, $\underset{^\Delta}{\epsilon}$$^{\delta+}_{ll}$, $\underset{^\Delta}{\epsilon}$$^{\delta+}_{tt}$, and $\underset{^\Delta}{\epsilon}$$^{\delta+}_{lt}$, and $\underset{^\Delta}{\epsilon}$$^{\delta-}_{ll}$, $\underset{^\Delta}{\epsilon}$$^{\delta-}_{tt}$, and $\underset{^\Delta}{\epsilon}$$^{\delta-}_{lt}$ in Figure \ref{fig:fiber_trend}. The colored bands represent the average and standard deviation. The gray bands represent the average strain changes during drying from 90 to 30\% RH (\textit{hygro}-expansion experiments) considering cycles 1$-$2 for all fibers (i.e. $\underset{^\Delta}{\bar{\epsilon}}$$^{\gamma-}_{ll}$, $\underset{^\Delta}{\bar{\epsilon}}$$^{\gamma-}_{tt}$, and $\underset{^\Delta}{\bar{\epsilon}}$$^{\gamma-}_{lt}$, given in Table \ref{tab:fiber_hygro}), including their standard deviations.}
	\label{fig:fiber_hydro}
\end{figure}
For quantitative comparison, the maximum swelling or shrinkage during the wetting cycle, respectively, $\underset{^\Delta}{\epsilon}$$^{\delta+}_{ll}$, $\underset{^\Delta}{\epsilon}$$^{\delta+}_{tt}$, and $\underset{^\Delta}{\epsilon}$$^{\delta+}_{lt}$, and $\underset{^\Delta}{\epsilon}$$^{\delta-}_{ll}$, $\underset{^\Delta}{\epsilon}$$^{\delta-}_{tt}$, and $\underset{^\Delta}{\epsilon}$$^{\delta-}_{lt}$, as annotated in Figure \ref{fig:fiber_trend}, are extracted for every FD and RD, HW and SW fiber and separately displayed along with their averages and standard deviations in Figure \ref{fig:fiber_hydro}. The corresponding values of the fiber \textit{hydro}-expansion are also given in Table \ref{tab:fiber_hydro}. The fiber shrinkage during the \textit{hygro}-expansion experiments, i.e. $\underset{^\Delta}{\bar{\epsilon}}$$^{\gamma-}_{ll}$, $\underset{^\Delta}{\bar{\epsilon}}$$^{\gamma-}_{tt}$, and $\underset{^\Delta}{\bar{\epsilon}}$$^{\gamma-}_{lt}$ are added, displaying that all fiber strain component magnitudes during the wetting cycle are significantly larger. Furthermore, the average longitudinal shrinkage during drying ($\underset{^\Delta}{\bar{\epsilon}}$$^{\delta-}_{ll}$) of the (HW and SW) RD fibers is larger than the average swelling during wetting ($\underset{^\Delta}{\bar{\epsilon}}$$^{\delta+}_{ll}$), hence the fiber is shorter after the wetting cycle. This is a logical result, as most fibers were dried under a tensile load, which is released during the wetting cycle. Additionally, this difference complies well with the negative release of longitudinal dried-in strain of the same fibers found in \citep{vonk2023res}. Finally, similar to the \textit{hygro}-expansion given in Figure \ref{fig:fiber_sheet_hygro}, the SW fibers display a significantly larger scatter in strain under \textit{hydro}-expansion than the HW fibers, and, similar as for \textit{hygro}-expansion (Figure \ref{fig:fiber_sheet_hygro}), $\underset{^\Delta}{\bar{\epsilon}}$$^{\delta}_{lt}$ of the FD and RD SW fibers remains around zero. \\ \indent
To relate the data to existing works, \cite{tydeman1966transverse} showed that eleven SW fibers, on average, shrunk 18.36$\pm$6.70\% in transverse direction from full wet to dry, and \cite{page1963transverse} showed that a SW fiber shrunk 16\% from wet to dry (20\% RH), both complying well with the transverse shrinkage of 15.4$\pm$2.4\% and 13.4$\pm$2.5\% found for, respectively, the FD and RD SW fibers here. The presented results are, however, different from the transverse shrinkage of around 40\% found by \cite{weise1995changes}. \\ \indent
To study the evolution of the fiber strain during \textit{hydro}-expansion, the transient \textit{hydro}-expansion during the wetting cycle (excluding the drying period) of four FD and four RD SW fibers is given in Figure \ref{fig:fiber_trend_FD_RD}. Firstly, the step in strain which stabilizes over time observed during the sheet-scale \textit{hydro}-expansion, is also visible in the transient \textit{hydro}-expansion of all SW fibers during the wetting cycle (more specifically, during the wetting period, as annotated in Figure \ref{fig:fiber_trend} (a)). Note that for the fiber experiments it is more difficult to reliably state that this strain step cannot be attributed to the transition between \textit{hygro}- and \textit{hydro}-expansion, unlike for to the sheet-scale \textit{hydro}-expansion experiments. However, small condensation droplets were visible near all fibers before the strain step initiated, suggesting that the fibers were already exhibiting \textit{hydro}-expansion, and a different mechanism drives the observed strain steps, similar to the handsheet \textit{hydro}-expansion observations.\\ \indent
\textbf{\begin{figure}[t!]
		\centering
		\includegraphics[width=0.5\textwidth]{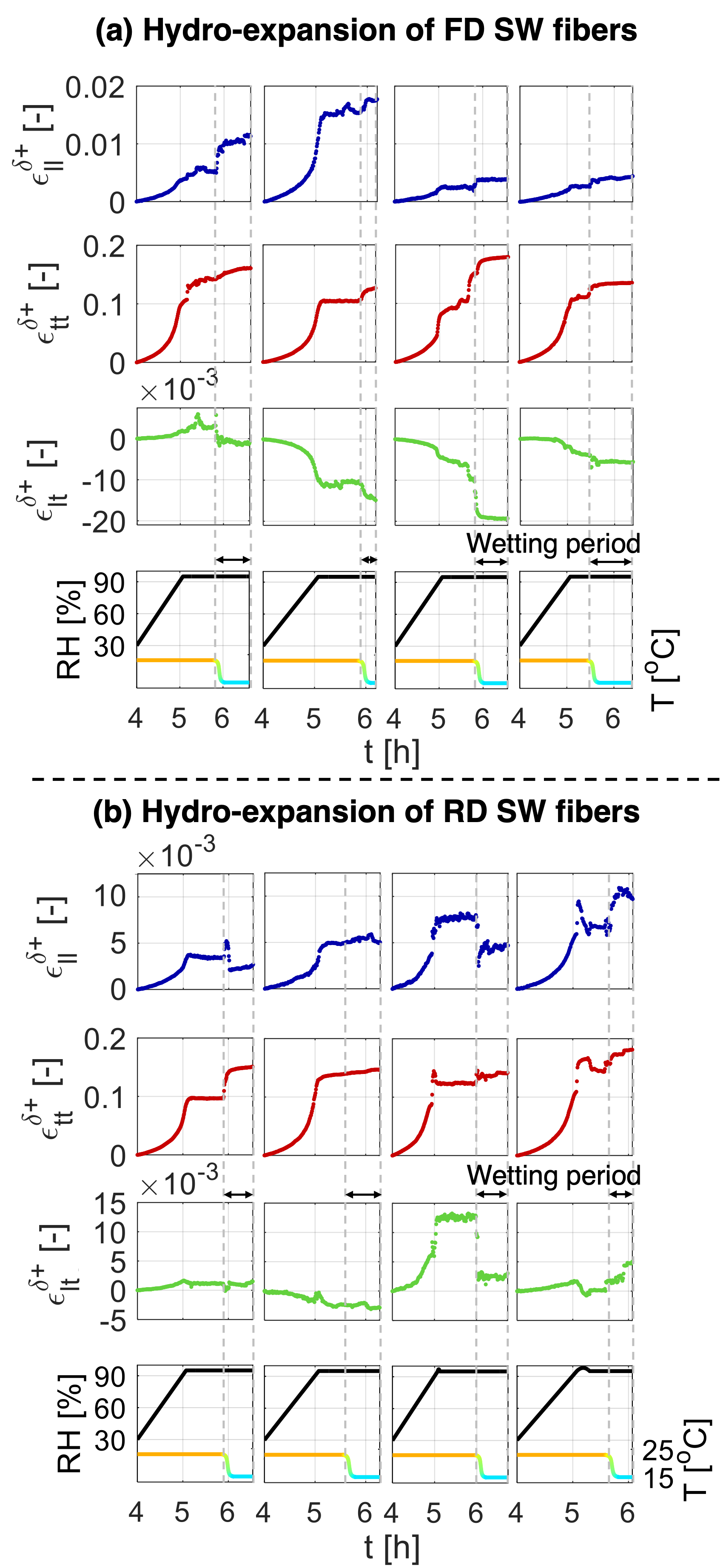}
		\caption{The longitudinal, transverse, and shear strain evolution of (a) four (out of ten) FD and (b) RD SW fibers during the wetting cycle (excluding the drying period), i.e., respectively, $\epsilon^{\delta+}_{ll}$, $\epsilon^{\delta+}_{tt}$, and $\epsilon^{\delta+}_{lt}$. The curves are vertically shifted to start at zero strain for better comparison between the fibers. The RH of the third and fourth RD SW fiber revealed an overshoot to 98\% RH when the set-point was supposed to be 95\% RH. Consequently, droplet formation was already visible and the fiber exhibited \textit{hydro}-expansion, hence the peaks in the strain curves.}
		\label{fig:fiber_trend_FD_RD}
\end{figure}}
Moreover, the sheet-scale \textit{hydro}-expansion stabilizes much slower than the fibers. This can be explained by the handsheet \textit{hydro}-expansion curve representing the response of all fibers constituting the sheet which are required to equilibrate before the handsheet is in full equilibrium, whereas this is not the case for single fibers. Furthermore, the RD SW fibers display both small positive and small negative steps in $\epsilon^{\delta+}_{ll}$, while the FD SW fibers only display (larger) positive steps in $\epsilon^{\delta+}_{ll}$, indicating that some of the RD SW fibers shorten during full wetting (in a freely manner). The amplitude ($A$) and saturation half-times of the strain steps ($\tau_{1/2}$) of all tested SW fibers are obtained by curve fitting the strain data during the wetting period as displayed in Figure \ref{fig:timescales}. Figure \ref{fig:timescales} provides $A$ and $\tau_{1/2}$ of each fiber separately (black open circles), together with the average values of the ensemble including the standard deviation of the mean (colored solid circles with error bars). Interestingly, for both $A$ and $\tau_{1/2}$, the standard deviations of the mean do not overlap, indicating that the three fiber strain directions have a significantly different $A$ and $\tau_{1/2}$. The fibers exhibit a more instantaneous step (lower $\tau_{1/2}$) in $\epsilon^{\delta+}_{ll}$, followed by a gradual increase (higher $\tau_{1/2}$) of $\epsilon^{\delta+}_{tt}$ over time, which is reflected at the sheet scale by a jump and subsequent gradual saturation of the \textit{hydro}-expansion $\epsilon^{\delta+}_s$ (Figure \ref{fig:sheet_hydro}). In order to unravel this strain step, which is present in both the transient fiber and sheet \textit{hydro}-expansion, a mechanism-based hypothesis is formulated next. 

\subsection{Hypothesis to explain the transient fiber and sheet hydro-expansion}
\begin{figure}[t!]
	\centering
	\includegraphics[width=0.75\textwidth,trim=6 2 3 3,clip]{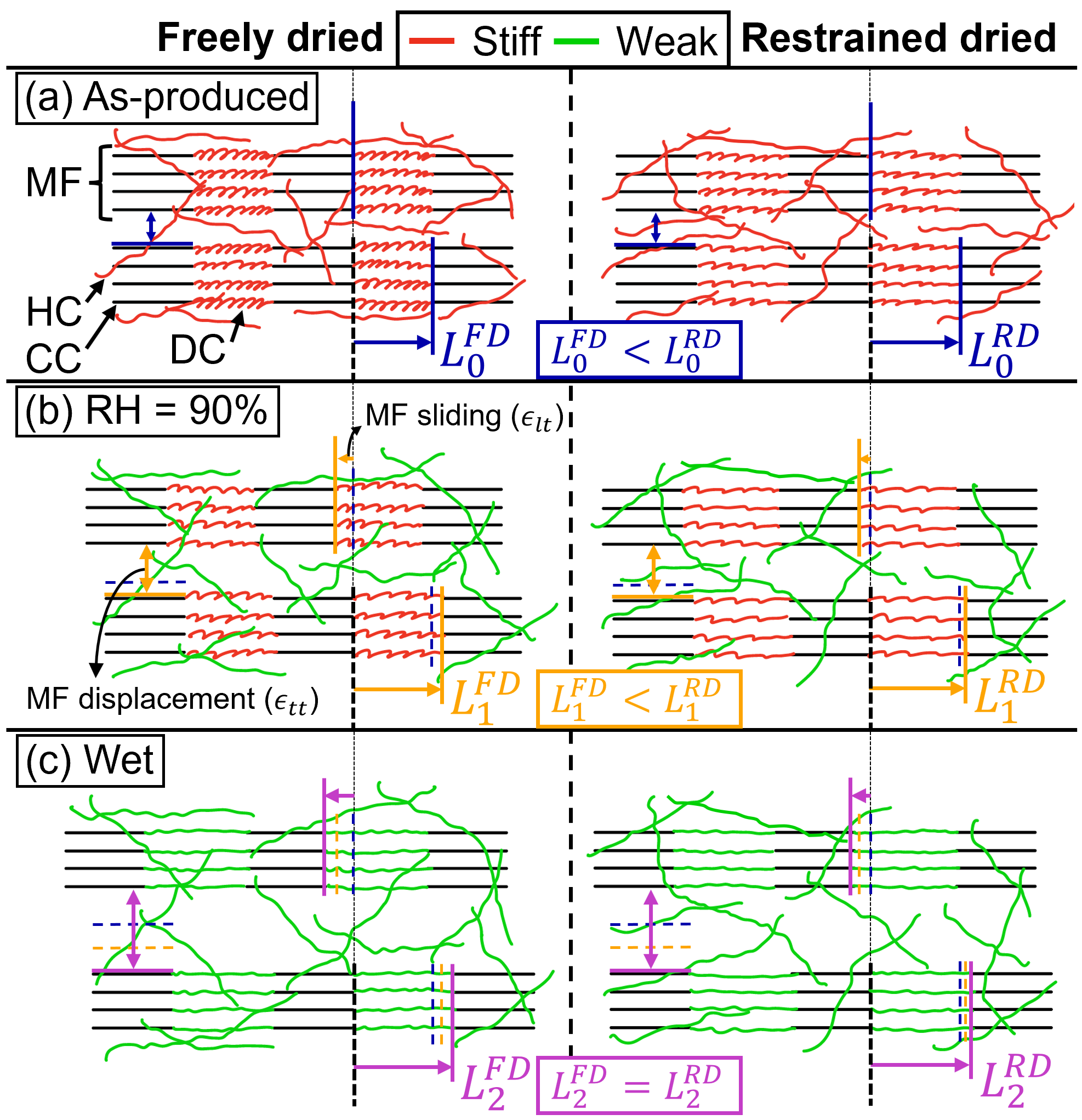}
	\caption{Schematic representation of the fiber wall structure in as-produced state, 90\% RH state, and fully wet state, to support the proposed hypothesis to explain the FD and RD, sheet and fiber, \textit{hygro}-expansion and \textit{hydro}-expansion findings. $L$ = length of the "dislocated cellulose regions". MF = micro-fibril, HC = hemi-cellulose, CC = crystalline cellulose, and DC = "dislocated cellulose".}
	\label{fig:hypot}
\end{figure}

To summarize the key observations so far, $\epsilon^{\delta+}_s$ and $MC^{\delta+}_s$ during the sheet-scale \textit{hydro}-expansion response displayed a sudden step, which was also visible for the fiber-scale \textit{hydro}-expansion. The step during the sheet-scale \textit{hydro}-expansion was significantly larger for FD compared to RD. Similarly, the SW fibers exhibited and average step in $\epsilon^{\delta+}_{ll}$ which was larger for FD than RD. Finally, the SW fibers exhibit distinctly different longitudinal, transverse, and shear strain step magnitudes and saturation half-times during the wetting period. \\ \indent
In order to explain these observations, a comprehensive hypothesis is formulated which is speculative but consistent. Since both the handsheets and SW fibers display a strain step, the driving mechanism must probably reside within the fibers rather than, e.g., the inter-fiber bonds. As generally known, the fiber cell wall consists of stiff cellulose micro-fibrils, consisting of alternating crystalline and so-called "dislocated regions" (both constituting \textsuperscript{$\sim$}50\% of the full micro-fibril), embedded inside an amorphous hemi-cellulose matrix \citep{agarwal2013estimation}. The hypothesized swelling mechanisms of a paper fiber (Figure \ref{fig:fiber_sheet_hygro}) from dry (as-produced) to 90\% RH is schematically depicted for FD and RD in Figure \ref{fig:hypot}, with the as-produced FD and RD fiber wall structure shown in (a) and the response after \textit{hygro}- and \textit{hydro}-expansion shown in, respectively, (b) and (c). \\ \indent
Starting with the \textit{hygro}-expansion, during water uptake, the hemi-cellulose (with a high porosity) absorbs most of the water and softens around 60\% RH, whereas the "dislocated cellulose regions" inside the micro-fibrils (having a low porosity) only slightly swell, see \citep{paajanen2022nanoscale}. Because the micro-fibril angle (MFA) of SW fibers is relatively low \citep{barnett2004cellulose, cown2004wood}, swelling of the hemi-cellulose results in relatively large transverse displacements and minor sliding of the micro-fibrils, yielding the $\underset{^\Delta}{\bar{\epsilon}}$$^{\gamma+}_{tt}$ and $\underset{^\Delta}{\bar{\epsilon}}$$^{\gamma+}_{lt}$ values of the fibers, whereas swelling of the "dislocated cellulose regions" is the main cause for the increase of $\underset{^\Delta}{\bar{\epsilon}}$$^{\gamma+}_{ll}$ of the fibers (Figure \ref{fig:fiber_sheet_hygro} (a)). Please note that Figure \ref{fig:fiber_sheet_hygro} displays the fiber strain changes during drying of RH cycles 1$-$2, which are similar to wetting (excluding the wetting slope of cycle 1), as, shown in Figure \ref{fig:fiber_trend}. Furthermore, the swelling of the "dislocated cellulose" is larger for FD than RD (i.e. $\frac{L_1^{FD}-L_0^{FD}}{L_0^{FD}}>\frac{L_1^{RD}-L_0^{RD}}{L_0^{RD}}$), due to the alignment of the regions as a result of restrained drying as discussed in the introduction, hence the lower $\underset{^\Delta}{\bar{\epsilon}}$$^{\gamma-}_{ll}$ in Figure \ref{fig:fiber_sheet_hygro} (a). Because larger $MC$ levels are attained during the fiber and handsheet \textit{hydro}-expansion experiments compared to the 90\% RH \textit{hygro}-expansion measurements, this hypothesized swelling mechanism is extended to describe the step in strain and $MC$ observed during the \textit{hydro}-expansion of FD and RD, HW and SW handsheets and fibers (Figures \ref{fig:sheet_hydro} and \ref{fig:fiber_trend_FD_RD}).\\ \indent
First, a hypothesis is proposed to explain the strain evolution of the handsheets and SW fibers during \textit{hydro}-expansion, for the moment disregarding the type of drying procedure (FD or RD). For the handsheets, the step in $\epsilon^{\delta+}_s$ and $MC^{\delta+}_s$ occurs after the $MC^{\delta+}_s$ reaches a specific critical value (of around 20\%), suggesting that the material exhibits a phase transformation, e.g. from (semi-)crystalline to amorphous. Note that this $MC$ level is far above the 90\% RH $MC$ level. The already amorphous hemi-cellulose is unlikely to transform due to the higher $MC$ level, whereas the crystalline cellulose regions remain stable, even at higher $MC$ levels. Hence, the so-called "dislocated regions" in the cellulose micro-fibrils are expected to soften through a phase transformation \citep{marchessault1957experimental, salmen1982temperature, salmen1987development, agarwal2018new}. It is presumed that at a certain $MC$ level, a critical capillary pressure is reached inside the fiber that is sufficiently high to "force" water molecules into the weak secondary bonds (hydrogen bonds) which are holding together the polymer chains in the "dislocated cellulose regions", thereby unlocking the inter-molecular bonds. This in turn results in softening of the "dislocated cellulose regions", entailing a sudden increase in water uptake in the now sparser molecular network. \\ \indent
Hence, considering the SW fibers, due to their low MFA, the process of softening of the "dislocated regions" in the cellulose micro-fibrils would mainly affect $\epsilon^{\delta+}_{ll}$, and to a lesser extent $\epsilon^{\delta+}_{tt}$ and $\epsilon^{\delta+}_{lt}$. In contrast, however, both $\epsilon^{\delta+}_{tt}$ and $\epsilon^{\delta+}_{lt}$ show steps that are significantly larger in magnitude and saturate slower than the step in $\epsilon^{\delta+}_{ll}$, as displayed by Figure \ref{fig:timescales}. This indicates the presence of a second mechanism that becomes active when the "dislocated cellulose regions" soften. To find an explanation for the different characteristics of the $\epsilon^{\delta+}_{ll}$, $\epsilon^{\delta+}_{tt}$, and $\epsilon^{\delta+}_{lt}$ curves, please consider that $\epsilon^{\delta+}_{tt}$ and $\epsilon^{\delta+}_{lt}$ are mainly driven by hemi-cellulose swelling and water residing between the micro-fibrils (pores), whereas $\epsilon^{\delta+}_{ll}$ is driven by swelling of the "dislocated cellulose regions" \citep{paajanen2022nanoscale}. Until the "dislocated cellulose regions" soften (leading to a jump in $\epsilon^{\delta+}_{ll}$, which is a fast process with a small $\tau_{1/2}$ of $\epsilon^{\delta+}_{ll}$ in Figure \ref{fig:timescales}), the hemi-cellulose ($\epsilon^{\delta+}_{tt}$ and $\epsilon^{\delta+}_{lt}$) was likely prevented from further swelling to its equilibrium volume, thus building up internal stress (pressure), which after softening of the "dislocated cellulose regions" starts to relax (gradually) through the uptake of more water (explaining the higher $\tau_{1/2}$). Hence, the hemi-cellulose was restrained from further swelling until unlocking and softening of the "dislocated cellulose regions". Probably candidates for this locking mechanism are (i) the helical micro-fibril structure in the fiber wall, as this requires the micro-fibrils to extend to enable the fiber to widen and/or lengthen; (ii) the interactions between the different S layers constituting the fiber wall, as kinematic compatibility of the S layers during swelling without building up residual stress is only feasible when the extension in the micro-fibril direction is not blocked. Furthermore, the smaller half-time of $\epsilon^{\delta+}_{lt}$ compared to $\epsilon^{\delta+}_{tt}$ in Figure \ref{fig:timescales} may indicate that the sliding of the parallel micro-fibrils saturates faster after softening of the "dislocated cellulose regions" has completed, compared to the relative (perpendicular) movement of the fibrils accommodating $\epsilon^{\delta+}_{tt}$, which would be logical because softening of the "dislocated cellulose" mainly affects the longitudinal (parallel) micro-fibril direction. \\ \indent
Next, the effect of the drying procedure is incorporated into the hypothesis to explain the larger step in the transient $\epsilon^{\delta+}_s$ and $MC^{\delta+}_s$ during wetting of FD compared to RD handsheets, which also occurs later for FD than RD. Due to the alignment and stretching of, respectively, the cellulose micro-fibrils and the "dislocated cellulose regions" inside the micro-fibrils as a result of restrained drying, as represented in Figure \ref{fig:hypot} (a), the "dislocated cellulose regions" in the RD fibers can stretch less (i.e. $\frac{L_2^{FD}-L_0^{FD}}{L_0^{FD}}>\frac{L_2^{RD}-L_0^{RD}}{L_0^{RD}}$) or they can even shrink when softening sets in, see Figure \ref{fig:fiber_hydro} (b) \citep{salmen1987development, khodayari2020tensile}. This explains the smaller step in the transient $\epsilon^{\delta+}_s$ for RD compared to FD handsheets, which is supported by the smaller average step in transient $\epsilon^{\delta+}_{ll}$ of the RD fibers compared to the FD SW fibers. Finally, after sufficiently long wetting under free boundary conditions, both the FD and RD fibers should "transform" to the same cell wall structure, which is shown in Figure \ref{fig:hypot} (c) with the same expected length of the (saturated) "dislocated cellulose regions" (i.e. $L^{FD}_2 = L^{RD}_2$). In practice, however, as shown in \citep{vonk2023res}, RD SW fibers are not able to completely transform into FD fibers even after maintaining the SW fibers for 2 days in water, which was attributed to the fact that SW fibers typically have a partially collapsed lumen (due to their large diameter and small cell wall thickness), which prohibits rearrangement of the helical fiber structure. Even with this additional complexity, it can be concluded that all SW fiber and sheet-scale observations can be reasonably well explained with the proposed hypothesis of the softening of the "dislocated cellulose regions" unlocking further hemi-cellulose swelling. Moreover, the proposed hypothesis should equally hold for HW fibers. It is, however, less fruitful to apply this theory on the HW results, because the transient \textit{hydro}-expansion response of the HW fibers is dictated by the slow introduction of water in the HW fiber experiments (see Figure \ref{fig:fiber_method}), which explains why no step in the strain components (with different timescales) is observed for the HW fibers. Nevertheless, the hypothesis still explains the qualitative curves observed in the HW fibers.
 
\subsection{Sheet-scale shrinkage predicted by the inter-fiber bond model}
\begin{figure}[t!]
	\centering
	\includegraphics[width=0.5\textwidth]{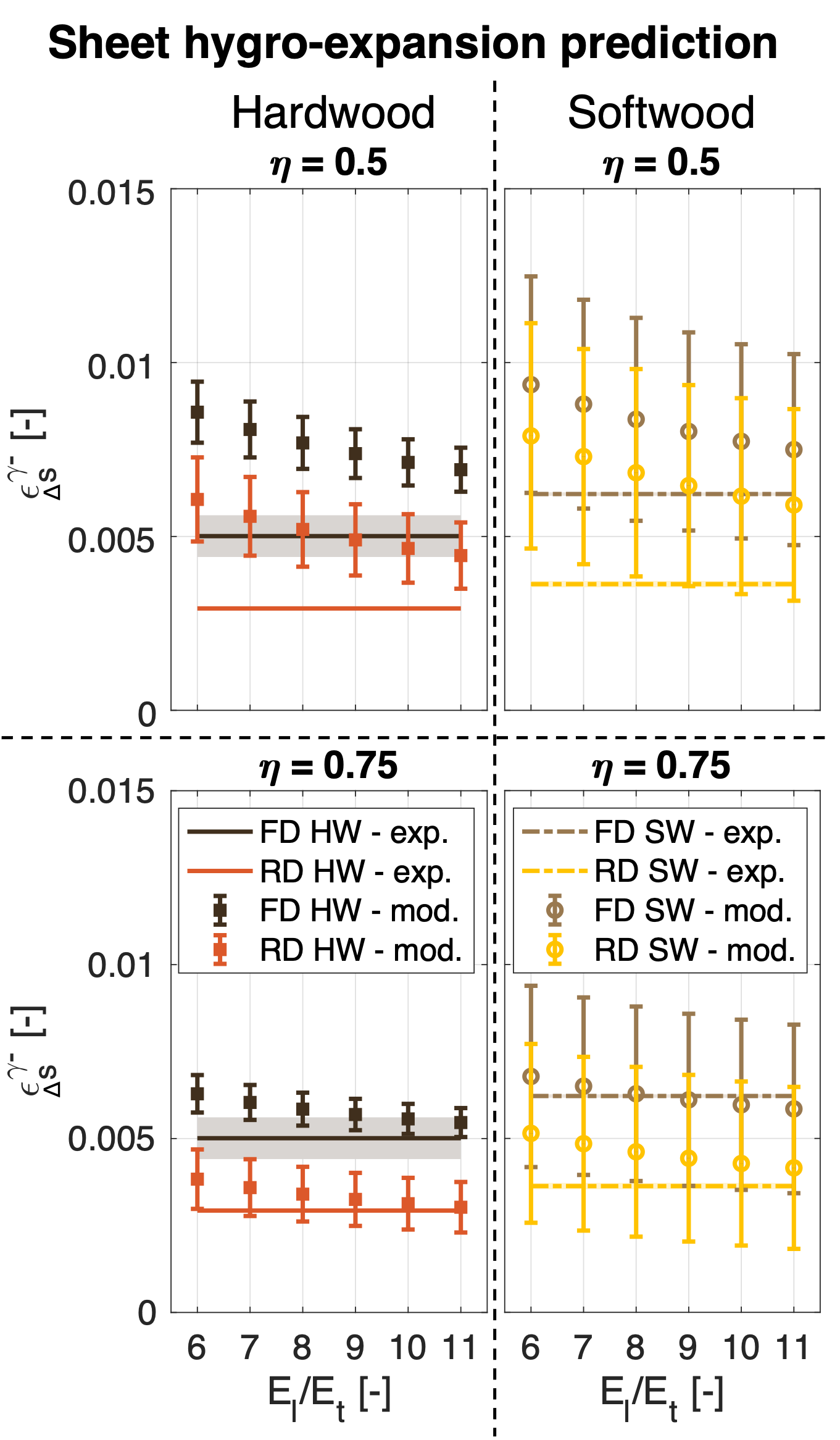}
	\caption{Comparison between the sheet-scale shrinkage during \textit{hygro}-expansion ($\underset{^\Delta}{\epsilon}$$^{\gamma-}_{s}$) predicted by the analytical model, using the fiber shrinkage characteristics during \textit{hygro}-expansion, i.e. $\underset{^\Delta}{\bar{\epsilon}}$$^{\gamma-}_{ll}$ and $\underset{^\Delta}{\bar{\epsilon}}$$^{\gamma-}_{tt}$ given in Figure \ref{fig:fiber_sheet_hygro}, and the experimental sheet-scale shrinkage ($\underset{^\Delta}{\bar{\epsilon}}$$^{\gamma-}_{s}$ given in Figure \ref{fig:fiber_sheet_hygro} (a)). The effect of the fibers' longitudinal to transverse stiffness ratio ($E_l/E_t$) and the free fiber length ($\eta$) is studied.}
\label{fig:sheet_prediction_hygro}
\end{figure}

The sheet-scale shrinkage during \textit{hygro}-expansion ($\underset{^\Delta}{\epsilon}$$^{\gamma-}_{s}$) predicted by the inter-fiber bond model, using the fiber shrinkage characteristics during \textit{hygro}-expansion, i.e. $\underset{^\Delta}{\bar{\epsilon}}$$^{\gamma-}_{ll}$ and $\underset{^\Delta}{\bar{\epsilon}}$$^{\gamma-}_{tt}$ displayed in Figure \ref{fig:fiber_sheet_hygro}, versus the transverse to longitudinal fiber stiffness ratio ($E_t/E_l$) for FD and RD, HW and SW is given in Figure \ref{fig:sheet_prediction_hygro}. The horizontal bands represent the experimentally obtained handsheet shrinkage ($\underset{^\Delta}{\bar{\epsilon}}$$^{\gamma-}_{s}$) given in Figure \ref{fig:fiber_sheet_hygro} (a) including standard deviations.\\ \indent
Let us first only consider the results predicted by the model with a free fiber length ($\eta$) of the 0.5. The predicted sheet strain reduces when the longitudinal to transverse stiffness ratio ($E_l/E_t$) increases, which makes sense because the $\epsilon^{\gamma-}_{tt}$ contribution in the bonded area becomes less due to the relatively larger $E_l$ of the bonded fiber. The model also correctly predicts that $\underset{^\Delta}{\epsilon}$$^{\gamma-}_{s}$ of the FD sheets is larger than the RD sheets, and the SW larger than the HW, similar to the experiments. FD being larger than RD is mainly driven by the larger $\epsilon^{\gamma-}_{ll}$ being larger for FD, because $\epsilon^{\gamma-}_{tt}$ is almost equal for FD and RD. So the model predicts the correct trends. Quantitatively, however, the model clearly overpredicts $\underset{^\Delta}{\epsilon}$$^{\gamma-}_{s}$, with the prediction only coming close to the experimentally obtained $\underset{^\Delta}{\epsilon}$$^{\gamma-}_{s}$ for larger fiber stiffness ratios and only for SW.\\ \indent
One may argue, however, that the fiber bonds are actually only bonded for \textsuperscript{$\sim$}50\%, e.g. \cite{sormunen2019x} showed using X-ray computed tomography with a high spatial resolution of 128 nm that the bonded area of various inter-fiber bonds (custom made with high pressures) is in reality around 60\%. Earlier X-ray CT measurements \citep{wernersson2014characterisations, borodulina2016extracting, urstoger2020microstructure} were performed with a spatial resolution of only \textsuperscript{$\sim$}0.5-1 $\mu$m, therefore, the measured bonded area may have been over-predicted. The effect of \textsuperscript{$\sim$}50\% bonding of the inter-fiber bonds may be approximated by assuming that the free fiber length is around 75\% (i.e. $\eta$ equals 0.75), resulting in a significantly better match with the experiments, see Figure \ref{fig:sheet_prediction_hygro}. Additionally, $\eta$ equals 0.75 is confirmed by \cite{torgnysdotter2007link} who found a fiber contact area of \textsuperscript{$\sim$}30\% for non-surface-charged paper. Note that the transverse strain transfer is equal for bonds which are fully bonded or only bonded at the edges, because the transverse strain transfer is mostly carried out by the outer edges of the bonded area. hence, in reality, a larger $eta$ can still result in the same sheet expansion as a lower $\eta$, which is not incorporated in the model. Furthermore, due to the considered orthogonal bond, the here-proposed model always predicts an overestimate of the actual sheet-scale shrinkage, as an additional diagonal fiber over the bonded area (as used in \citep{bosco2015predicting}) would not change $\eta$, while it lowers $\epsilon^{\gamma-}_b$. \\ \indent
The model has also been assessed for predicting the sheet-scale \textit{hydro}-expansion ($\underset{^\Delta}{\epsilon}$$^{\delta}_{s}$) during wetting and drying using the fiber characteristics given in Figure \ref{fig:fiber_hydro}. However, for both the 50 and 75\% free fiber length, the model significantly under-predicts the experimentally obtained $\underset{^\Delta}{\epsilon}$$^{\delta}_{s}$ given in Table \ref{tab:exp_sheet_hydro}. This holds especially for the FD sheets, which reveal a much larger experimental $\underset{^\Delta}{\epsilon}$$^{\delta}_{s}$ compared to the longitudinal fiber \textit{hydro}-expansion. Possible explanations for this discrepancy may be the increased transverse strain contribution to the sheet scale during \textit{hydro}-expansion. Furthermore, inter-fiber bonds may partially debond and rebond during the \textit{hydro}-expansion experiments, hence changing the 3D geometry of network, whereas this is unlikely to occur in the \textit{hygro}-expansion experiments. Finally, non-linear behavior such as plasticity is likely to occur at such high $MC$ levels. Clearly, these effects are not incorporated into the simple model.\\ \indent
Even though the model is unable to predict the sheet-scale \textit{hydro}-expansion of paper sheets, the model adequately matches the experiments for \textit{hygro}-expansion, at least qualitatively. This demonstrates that such a (simple) analytical model is useful for identifying the ingredients governing the \textit{hygro}-expansion behavior. 

\section{Conclusion}
The \textit{hygro}- and \textit{hydro}-expansion of freely and restrained dried handsheets and fibers extracted from these handsheets have been investigated, allowing to explore the relevant mechanisms over time along with the accompanying dimensional changes. Previous works have suggested that the dimensional stability of a paper sheet is strongly affected by the stress applied during drying from a saturated state (due to liquid water, i.e. \textit{hydro}-expansion) to lower moisture content ($MC$) levels, i.e. at 90\% relative humidity (RH) (\textit{hygro}-expansion). Therefore \textit{hygro}-expansion measurements were conducted by changing the RH around fibers and handsheets, while \textit{hydro}-expansion is realized by cooling the specimen to trigger condensation at the fibers and handsheets. To monitor the sheet \textit{hydro}-expansion, a novel method was developed. During all experiments, the dimensional changes are determined using state-of-the-art digital image correlation techniques, enabling the continuous transient full-field (longitudinal, transverse and shear) fiber, and (isotropic) in-plane sheet \textit{hygro}- and \textit{hydro}-expansion. Additionally, the transient $MC$ has also been monitored during the sheet \textit{hydro}-expansion measurements. Furthermore, the experimental fiber and sheet data enables to assess a simple inter-fiber bond model in order to predict the sheet expansion from the experimentally obtained fiber characteristics for various ratios between the longitudinal and transverse fiber stiffness. \\ \indent
For \textit{hygro}- and \textit{hydro}-expansion, the freely dried handsheets shrink more than the restrained dried, and the softwood handsheets shrink more than the hardwood. Furthermore, the longitudinal fiber \textit{hygro}- and \textit{hydro}-expansion is distinctly lower for restrained drying, explaining the sheet-scale differences, because it was found that the transverse strain contribution to the sheet scale is relatively low. The transient sheet-scale \textit{hydro}-expansion experiments reveal a sudden step in strain and $MC$, which was also visible in the transient \textit{hydro}-expansion of freely dried softwood fibers. This step was attributed to the capillary pressure inside the fiber reaching a critical level, "forcing" the water into the so-called "dislocated cellulose regions" inside the micro-fibrils, which consequently soften, enabling the fiber to continue swelling. During the handsheet \textit{hydro}-expansion experiments, the strain step is significantly larger and occurs at higher $MC$ levels for freely dried compared to restrained dried handsheets. The difference in magnitude of the strain step is attributed to the "dislocated regions" in the cellulose micro-fibrils being stretched after restrained drying, therefore not being able to stretch upon softening, in contrast to their freely dried counterparts. This hypothesis is strongly supported by the restrained dried softwood fibers which reveal both swelling and shrinkage in longitudinal direction when the softening is expected to occur, whereas the freely dried fibers only show swelling, explaining the lower resulting strain step for the restrained dried handsheets. \\ \indent
The inter-fiber bond model qualitatively predicts the sheet \textit{hygro}-expansion trends, i.e. the larger \textit{hygro}-expansion of (i) FD compared to RD handsheets, and (ii) SW compared to HW handsheets. A decreasing trend in sheet expansion for an increasing longitudinal to transverse fiber stiffness ratio is predicted by the inter-fiber bond model. Furthermore, the model overestimates the sheet-scale \textit{hygro}-expansion for a free fiber length of 0.5. A free fiber length of 0.75 was also assessed, which complies better with the literature, resulting in a better quantitative match between the model and the experiments for \textit{hygro}-expansion. Regarding \textit{hydro}-expansion, the model underpredicts the experiments, which could be due to the simplicity of the model missing some key features affecting \textit{hydro}-expansion, e.g. breaking and reformation of bonds.

\appendix \newpage
\setcounter{figure}{0} 
\setcounter{table}{0} 
\setcounter{page}{1}

\section{Supplementary material}

\subsection{Experimental overview}
\label{app:experimental}
\begin{table}[H]
	\centering
	\caption{Total amount of samples tested samples which are the same for \textit{hygro}- and \textit{hydro}-expansion.}
	\label{tab:samples}
	\resizebox{0.25\textwidth}{!}{
		\begin{tabular}{lcc}
			\toprule
			\textbf{FD} & sheets & fibers \\ \midrule
			HW & 2     & 5      \\
			SW & 2     & 10     \\ \bottomrule \toprule
			\textbf{RD} & sheets & fibers \\ \midrule
			HW & 2     & 9  \\
			SW & 2     & 10  \\ \bottomrule    
	\end{tabular}}
\end{table}

\subsection{Sheet and fiber hygro-expansion}
\label{app:fiber_sheet_hygro}

\begin{table}[H]
	\centering
	\caption{Average longitudinal ($\underset{^\Delta}{\bar{\epsilon}}$$^{\gamma-}_{ll}$) and transverse shrinkage ($\underset{^\Delta}{\bar{\epsilon}}$$^{\gamma-}_{tt}$), and shear strain change ($\underset{^\Delta}{\bar{\epsilon}}$$^{\gamma-}_{lt}$) considering the drying slopes from 90 to 30\% RH of cycles 1$-$2 (\textit{hygro}-expansion) for five FD and nine RD HW fibers, and ten FD and RD SW fibers, including their standard deviation.}
	\label{tab:fiber_hygro}
	\resizebox{0.5\textwidth}{!}{
		\begin{tabular}{@{}lccc@{}}
			\toprule
			\textbf{FD} & $\underset{^\Delta}{\bar{\epsilon}}$$^{\gamma-}_{ll}$     [-]        & $\underset{^\Delta}{\bar{\epsilon}}$$^{\gamma-}_{tt}$ [-]   & $\underset{^\Delta}{\bar{\epsilon}}$$^{\gamma-}_{lt}$    [-]     \\ \midrule
			HW & 0.0040$\pm$0.0002 & 0.0627$\pm$0.0089 & -0.0077$\pm$0.0019 \\
			SW & 0.0042$\pm$0.0021 & 0.0705$\pm$0.0151& -0.0011$\pm$0.0033  \\ \bottomrule \toprule
			\textbf{RD} & $\underset{^\Delta}{\bar{\epsilon}}$$^{\gamma-}_{ll}$     [-]        & $\underset{^\Delta}{\bar{\epsilon}}$$^{\gamma-}_{tt}$ [-]   & $\underset{^\Delta}{\bar{\epsilon}}$$^{\gamma-}_{lt}$    [-]     \\ \midrule
			HW & 0.0016$\pm$0.0005 & 0.0589$\pm$0.0096 & -0.0034$\pm$0.0018\\
			SW & 0.0024$\pm$0.0019 & 0.0729$\pm$0.0191 & 0.0002$\pm$0.0024 \\ \bottomrule
	\end{tabular}}
\end{table}

\begin{table}[H]
	\centering
	\caption{The average shrinkage ($\underset{^\Delta}{\bar{\epsilon}}$$^{\gamma-}_{s}$ ) considering the six drying slopes from 90 to 30\% RH (\textit{hygro}-expansion) of two FD and RD, HW and SW handsheets, including their standard deviation.}
	\label{tab:exp_sheet_hygro}
	\resizebox{0.25\textwidth}{!}{
		\begin{tabular}{lc}
			\toprule
			\textbf{FD}    &  $\underset{^\Delta}{\bar{\epsilon}}$$^{\gamma-}_{s}$ [-]\\ \midrule
			HW    & 0.0050$\pm$0.0006 \\
			SW    & 0.0062$\pm$0.0001 \\ \bottomrule \toprule
			\textbf{RD}    &  $\underset{^\Delta}{\bar{\epsilon}}$$^{\gamma-}_{s}$ [-]\\ \midrule
			HW    & 0.0029$\pm$0.0001 \\
			SW    & 0.0036$\pm$0.0001\\ \bottomrule
	\end{tabular}}
\end{table}

\subsection{Sheet and fiber hydro-expansion}
\label{app:fiber_sheet_hydro}

\begin{table}[H]
	\centering
	\caption{The maximum swelling ($\underset{^\Delta}{\epsilon}$$^{\delta+}_{s}$) and shrinkage ($\underset{^\Delta}{\epsilon}$$^{\delta-}_{s}$) of a FD and RD, HW and SW handsheet during the sheet-scale \textit{hydro}-expansion experiments as annotated in Figure \ref{fig:sheet_hydro}.}
	\label{tab:exp_sheet_hydro}
	\resizebox{0.25\textwidth}{!}{
		\begin{tabular}{lcc}
			\toprule
			\textbf{FD}    &  $\underset{^\Delta}{\epsilon}$$^{\delta+}_{s}$ [-] & $\underset{^\Delta}{\epsilon}$$^{\delta-}_{s}$ [-]\\ \midrule
			HW    & 0.0173 & 0.0208\\
			SW    & 0.0192 & 0.0271\\ \bottomrule \toprule
			\textbf{RD}    &  $\underset{^\Delta}{\epsilon}$$^{\delta+}_{s}$ [-] & $\underset{^\Delta}{\epsilon}$$^{\delta-}_{s}$ [-]\\ \midrule
			HW    & 0.0056 & 0.0144\\
			SW    & 0.0102 & 0.0168\\ \bottomrule
	\end{tabular}}
\end{table}

\begin{table}[H]
	\centering
	\caption{Average longitudinal and transverse, and shear strain change for wetting and drying during the wetting cycle, i.e., respectively, $\underset{^\Delta}{\bar{\epsilon}}$$^{\delta+}_{ll}$, $\underset{^\Delta}{\bar{\epsilon}}$$^{\delta+}_{tt}$, and $\underset{^\Delta}{\bar{\epsilon}}$$^{\delta+}_{lt}$, and $\underset{^\Delta}{\bar{\epsilon}}$$^{\delta-}_{ll}$, $\underset{^\Delta}{\bar{\epsilon}}$$^{\delta-}_{tt}$, and $\underset{^\Delta}{\bar{\epsilon}}$$^{\delta-}_{lt}$ annotated in Figure \ref{fig:fiber_trend}, for five FD and nine RD HW fibers, and ten FD and RD SW fibers, including their standard deviation.}
	\label{tab:fiber_hydro}
	\resizebox{\textwidth}{!}{
		\begin{tabular}{@{}lccc|ccc@{}}
			\toprule
			\textbf{FD} & $\underset{^\Delta}{\bar{\epsilon}}$$^{\delta+}_{ll}$     [-]        & $\underset{^\Delta}{\bar{\epsilon}}$$^{\delta+}_{tt}$ [-]   & $\underset{^\Delta}{\bar{\epsilon}}$$^{\delta+}_{lt}$   [-]   & $\underset{^\Delta}{\bar{\epsilon}}$$^{\delta-}_{ll}$    [-]        & $\underset{^\Delta}{\bar{\epsilon}}$$^{\delta-}_{tt}$ [-]   & $\underset{^\Delta}{\bar{\epsilon}}$$^{\delta-}_{lt}$   [-]   \\ \midrule
			HW & 0.0064$\pm$0.0007 &0.0966$\pm$0.0099 & -0.0089$\pm$0.0011 &0.0068$\pm$0.0005 & 0.0998$\pm$0.0144 & -0.0103$\pm$0.0045 \\
			SW & 0.0086$\pm$0.0042 & 0.1441$\pm$0.0302& -0.0015$\pm$0.0097 & 0.0078$\pm$0.0052 & 0.1337$\pm$0.0254& -0.0026$\pm$0.0079\\ \bottomrule \toprule
			\textbf{RD} & $\underset{^\Delta}{\bar{\epsilon}}$$^{\delta+}_{ll}$     [-]        & $\underset{^\Delta}{\bar{\epsilon}}$$^{\delta+}_{tt}$ [-]   & $\underset{^\Delta}{\bar{\epsilon}}$$^{\delta+}_{lt}$   [-]   & $\underset{^\Delta}{\bar{\epsilon}}$$^{\delta-}_{ll}$    [-]        & $\underset{^\Delta}{\bar{\epsilon}}$$^{\delta-}_{tt}$ [-]   & $\underset{^\Delta}{\bar{\epsilon}}$$^{\delta-}_{lt}$   [-]   \\ \midrule
			HW & 0.0038$\pm$0.0012 & 0.0999$\pm$0.0352 & -0.0054$\pm$0.0045 & 0.0044$\pm$0.0014 & 0.0992$\pm$0.0365 & -0.0070$\pm$0.0034\\
			SW & 0.0058$\pm$0.0023 & 0.1629$\pm$0.0170 & -0.0004$\pm$0.0087 & 0.0077$\pm$0.0044 & 0.1542$\pm$0.0237 & -0.0025$\pm$0.0104\\ \bottomrule
	\end{tabular}}
\end{table}


\newpage
\begin{thebibliography}{}
	
	\bibitem[Agarwal et~al., 2018]{agarwal2018new}
	Agarwal, U.~P., Ralph, S.~A., Reiner, R.~S., and Baez, C. (2018).
	\newblock New cellulose crystallinity estimation method that differentiates
	between organized and crystalline phases.
	\newblock {\em Carbohydrate polymers}, 190:262--270.
	
	\bibitem[Agarwal et~al., 2013]{agarwal2013estimation}
	Agarwal, U.~P., Reiner, R.~R., and Ralph, S.~A. (2013).
	\newblock Estimation of cellulose crystallinity of lignocelluloses using
	near-ir ft-raman spectroscopy and comparison of the raman and segal-waxs
	methods.
	\newblock {\em Journal of agricultural and food chemistry}, 61(1):103--113.
	
	\bibitem[Alzweighi et~al., 2021]{alzweighi2021influence}
	Alzweighi, M., Mansour, R., Lahti, J., Hirn, U., and Kulachenko, A. (2021).
	\newblock The influence of structural variations on the constitutive response
	and strain variations in thin fibrous materials.
	\newblock {\em Acta Materialia}, 203:116460.
	
	\bibitem[Barnett and Bonham, 2004]{barnett2004cellulose}
	Barnett, J.~R. and Bonham, V.~A. (2004).
	\newblock Cellulose microfibril angle in the cell wall of wood fibres.
	\newblock {\em Biological reviews}, 79(2):461--472.
	
	\bibitem[Borodulina et~al., 2016]{borodulina2016extracting}
	Borodulina, S., Kulachenko, A., Wernersson, E. L.~G., and Luengo~Hendriks,
	C.~L. (2016).
	\newblock Extracting fiber and network connectivity data using microtomography
	images of paper.
	\newblock {\em Nordic Pulp \& Paper Research Journal}, 31(3):469--478.
	
	\bibitem[Bosco et~al., 2015a]{bosco2015bridging}
	Bosco, E., Bastawrous, M.~V., Peerlings, R. H.~J., Hoefnagels, J. P.~M., and
	Geers, M. G.~D. (2015a).
	\newblock Bridging network properties to the effective hygro-expansivity of
	paper: experiments and modelling.
	\newblock {\em Philosophical Magazine}, 95(28-30):3385--3401.
	
	\bibitem[Bosco et~al., 2015b]{bosco2015predicting}
	Bosco, E., Peerlings, R. H.~J., and Geers, M. G.~D. (2015b).
	\newblock Predicting hygro-elastic properties of paper sheets based on an
	idealized model of the underlying fibrous network.
	\newblock {\em International Journal of Solids and Structures}, 56:43--52.
	
	\bibitem[Bosco et~al., 2017a]{bosco2017asymptotic}
	Bosco, E., Peerlings, R. H.~J., and Geers, M. G.~D. (2017a).
	\newblock Asymptotic homogenization of hygro-thermo-mechanical properties of
	fibrous networks.
	\newblock {\em International Journal of Solids and Structures}, 115:180--189.
	
	\bibitem[Bosco et~al., 2017b]{bosco2017hygro}
	Bosco, E., Peerlings, R. H.~J., and Geers, M. G.~D. (2017b).
	\newblock Hygro-mechanical properties of paper fibrous networks through
	asymptotic homogenization and comparison with idealized models.
	\newblock {\em Mechanics of Materials}, 108:11--20.
	
	\bibitem[Brandberg et~al., 2020]{brandberg2020role}
	Brandberg, A., Motamedian, H.~R., Kulachenko, A., and Hirn, U. (2020).
	\newblock The role of the fiber and the bond in the hygroexpansion and curl of
	thin freely dried paper sheets.
	\newblock {\em International Journal of Solids and Structures}, 193:302--313.
	
	\bibitem[Cown et~al., 2004]{cown2004wood}
	Cown, D.~J., Ball, R.~D., and Riddell, M. J.~C. (2004).
	\newblock Wood density and microfibril angle in 10 pinus radiata clones:
	distribution and influence on product performance.
	\newblock {\em New Zealand Journal of Forestry Science}, 34(3):293.
	
	\bibitem[Czibula et~al., 2021]{czibula2021transverse}
	Czibula, C., Brandberg, A., Cordill, M.~J., Matkovi{\'c}, A., Glushko, O.,
	Czibula, C., Kulachenko, A., Teichert, C., and Hirn, U. (2021).
	\newblock The transverse and longitudinal elastic constants of pulp fibers in
	paper sheets.
	\newblock {\em Scientific reports}, 11(1):1--13.
	
	\bibitem[Fellers, 2007]{fellers2007interaction}
	Fellers, C. (2007).
	\newblock The interaction of paper with water vapour.
	\newblock {\em Paper Products--Physics and Technology}, pages 109--144.
	
	\bibitem[Jajcinovic et~al., 2018]{jajcinovic2018influence}
	Jajcinovic, M., Fischer, W.~J., Mautner, A., Bauer, W., and Hirn, U. (2018).
	\newblock Influence of relative humidity on the strength of hardwood and
	softwood pulp fibres and fibre to fibre joints.
	\newblock {\em Cellulose}, 25(4):2681--2690.
	
	\bibitem[Jentzen, 1964]{jentzen1964effect}
	Jentzen, C.~A. (1964).
	\newblock {\em The effect of stress applied during drying on some of the
		properties of individual pulp fibers}.
	\newblock PhD thesis, Georgia Institute of Technology.
	
	\bibitem[Joffre et~al., 2016]{joffre2016method}
	Joffre, T., Isaksson, P., Dumont, P. J.~J., Roscoat, S., Sticko, S.,
	Org{\'e}as, L., and Gamstedt, E.~K. (2016).
	\newblock A method to measure moisture induced swelling properties of a single
	wood cell.
	\newblock {\em Experimental Mechanics}, 56(5):723--733.
	
	\bibitem[Khodayari et~al., 2020]{khodayari2020tensile}
	Khodayari, A., Van~Vuure, A.~W., Hirn, U., and Seveno, D. (2020).
	\newblock Tensile behaviour of dislocated/crystalline cellulose fibrils at the
	nano scale.
	\newblock {\em Carbohydrate polymers}, 235:115946.
	
	\bibitem[Kulachenko et~al., 2005]{kulachenko2005tension}
	Kulachenko, A., Gradin, P., and Uesaka, T. (2005).
	\newblock Tension wrinkling and fluting in heatset web offset printing process.
	post buckling analyses.
	\newblock In {\em 13th Fundamental Research Symposium on Advances in Paper
		Science and Technology Location: Univ Cambridge, Cambridge, ENGLAND Date:
		SEP, 2005}, pages 1075--1099. The Pulp and Paper Fundamental Research
	Society.
	
	\bibitem[Larsson et~al., 2009a]{larsson2009influence}
	Larsson, P.~A., Hoc, M., and W{\aa}gberg, L. (2009a).
	\newblock The influence of grammage, moisture content, fibre furnish and
	chemical modifications on the hygro-and hydro-expansion of paper.
	\newblock In {\em 14th Fundamental Research Symposium on Advances in Pulp and
		Paper Research Location: St Annes Coll, Oxford, ENGLAND Date: SEP 13-18,
		2009}, pages 355--388.
	
	\bibitem[Larsson et~al., 2009b]{larsson2009novel}
	Larsson, P.~A., Hoc, M., and W{\aa}gberg, L. (2009b).
	\newblock A novel approach to study the hydroexpansion mechanisms of paper
	using spray technique.
	\newblock {\em Nordic Pulp \& Paper Research Journal}, 24(4):371--380.
	
	\bibitem[Larsson and W{\aa}gberg, 2008]{larsson2008influence}
	Larsson, P.~A. and W{\aa}gberg, L. (2008).
	\newblock Influence of fibre--fibre joint properties on the dimensional
	stability of paper.
	\newblock {\em Cellulose}, 15(4):515--525.
	
	\bibitem[Magnusson and {\"O}stlund, 2013]{magnusson2013numerical}
	Magnusson, M.~S. and {\"O}stlund, S. (2013).
	\newblock Numerical evaluation of interfibre joint strength measurements in
	terms of three-dimensional resultant forces and moments.
	\newblock {\em Cellulose}, 20(4):1691--1710.
	
	\bibitem[Marchessault and Howsmon, 1957]{marchessault1957experimental}
	Marchessault, R.~H. and Howsmon, J.~A. (1957).
	\newblock Experimental evaluation of the lateral-order distribution in
	cellulose.
	\newblock {\em Textile Research Journal}, 27(1):30--41.
	
	\bibitem[Meylan, 1972]{meylan1972influence}
	Meylan, B.~A. (1972).
	\newblock The influence of microfibril angle on the longitudinal
	shrinkage-moisture content relationship.
	\newblock {\em Wood science and technology}, 6(4):293--301.
	
	\bibitem[Motamedian and Kulachenko, 2019]{motamedian2019simulating}
	Motamedian, H.~R. and Kulachenko, A. (2019).
	\newblock Simulating the hygroexpansion of paper using a 3d beam network model
	and concurrent multiscale approach.
	\newblock {\em International Journal of Solids and Structures}, 161:23--41.
	
	\bibitem[Nanko and Wu, 1995]{nanko1995mechanisms}
	Nanko, H. and Wu, J. (1995).
	\newblock Mechanisms of paper shrinkage during drying.
	\newblock In {\em International paper physics conference, Niagara-on-the-Lake,
		Canada}, pages 103--113.
	
	\bibitem[Neggers et~al., 2016]{neggers2016image}
	Neggers, J., Blaysat, B., Hoefnagels, J. P.~M., and Geers, M. G.~D. (2016).
	\newblock On image gradients in digital image correlation.
	\newblock {\em International Journal for Numerical Methods in Engineering},
	105(4):243--260.
	
	\bibitem[Niskanen et~al., 1997]{niskanen1997dynamic}
	Niskanen, K.~J., Kuskowski, S.~J., and Bronkhorst, C.~A. (1997).
	\newblock Dynamic hygroexpansion of paperboards.
	\newblock {\em Nordic Pulp \& Paper Research Journal}, 12(2):103--110.
	
	\bibitem[Paajanen et~al., 2022]{paajanen2022nanoscale}
	Paajanen, A., Zitting, A., Rautkari, L., Ketoja, J.~A., and Penttil\"a, P.~A.
	(2022).
	\newblock Nanoscale mechanism of moisture-induced swelling in wood microfibril
	bundles.
	\newblock {\em Nano letters}, 22(13):5143--5150.
	
	\bibitem[Page and Tydeman, 1963]{page1963transverse}
	Page, D.~H. and Tydeman, P.~A. (1963).
	\newblock Transverse swelling and shrinkage of softwood tracheids.
	\newblock {\em Nature}, 199(4892):471--472.
	
	\bibitem[Pister and Dong, 1959]{pister1959elastic}
	Pister, K.~S. and Dong, S.~B. (1959).
	\newblock Elastic bending of layered plates.
	\newblock {\em Journal of the Engineering Mechanics Division}, 85(4):1--10.
	
	\bibitem[Reissner and Stavsky, 1961]{reissner1961bending}
	Reissner, E. and Stavsky, Y.~l. (1961).
	\newblock Bending and stretching of certain types of heterogeneous aeolotropic
	elastic plates.
	\newblock {\em Journal of Applied Mechanics}, 28(3):402--408.
	
	\bibitem[Salm{\'e}n et~al., 1985]{salmen1985plane}
	Salm{\'e}n, L., Fellers, C., and Htun, M. (1985).
	\newblock The in-plane and out-of-plane hygroexpansional properties of paper.
	\newblock {\em Papermaking Raw Materials}, 2:511--527.
	
	\bibitem[Salmén, 1982]{salmen1982temperature}
	Salmén, L. (1982).
	\newblock {\em Temperature and water induced softening behaviour of wood fiber
		based materials}.
	\newblock PhD thesis, Department of Paper Technology, Royal Institute of
	Technology Stockholm.
	
	\bibitem[Salmén et~al., 1987]{salmen1987development}
	Salmén, L., Fellers, C., and Htun, M. (1987).
	\newblock The development and release of dried-in stresses in paper.
	\newblock {\em Nordic Pulp \& Paper Research Journal}, 2(2):44--48.
	
	\bibitem[Samantray et~al., 2020]{samantray2020level}
	Samantray, P., Peerlings, R. H.~J., Bosco, E., Geers, M. G.~D., Massart, T.~J.,
	and Roko{\v{s}}, O. (2020).
	\newblock Level set-based extended finite element modeling of the response of
	fibrous networks under hygroscopic swelling.
	\newblock {\em Journal of Applied Mechanics}, 87(10).
	
	\bibitem[Shafqat and Hoefnagels, 2021]{shafqat2021cool}
	Shafqat, S. and Hoefnagels, J. P.~M. (2021).
	\newblock Cool, dry, nano-scale dic patterning of delicate, heterogeneous,
	non-planar specimens by micro-mist nebulization.
	\newblock {\em Experimental Mechanics}, 61(6):917--937.
	
	\bibitem[Sormunen et~al., 2019]{sormunen2019x}
	Sormunen, T., Ketola, A., Miettinen, A., Parkkonen, J., and Retulainen, E.
	(2019).
	\newblock X-ray nanotomography of individual pulp fibre bonds reveals the
	effect of wall thickness on contact area.
	\newblock {\em Scientific Reports}, 9(1):1--7.
	
	\bibitem[Torgnysdotter et~al., 2007]{torgnysdotter2007link}
	Torgnysdotter, A., Kulachenko, A., Gradin, P., and W{\aa}gberg, L. (2007).
	\newblock The link between the fiber contact zone and the physical properties
	of paper: a way to control paper properties.
	\newblock {\em Journal of composite materials}, 41(13):1619--1633.
	
	\bibitem[Tydeman et~al., 1966]{tydeman1966transverse}
	Tydeman, P.~A., Wembridge, D.~R., and Page, D.~H. (1966).
	\newblock Transverse shrinkage of individual fibres by micro-radiography.
	\newblock In {\em Consolidation of the paper web}, pages 119--144.
	
	\bibitem[Uesaka, 1994]{uesaka1994general}
	Uesaka, T. (1994).
	\newblock General formula for hygroexpansion of paper.
	\newblock {\em Journal of materials science}, 29(9):2373--2377.
	
	\bibitem[Uesaka et~al., 1992]{uesaka1992characterization}
	Uesaka, T., Moss, C., and Nanri, Y. (1992).
	\newblock The characterization of hygroexpansivity of paper.
	\newblock {\em Journal of pulp and paper science}, 18(1):J11--J16.
	
	\bibitem[Urst{\"o}ger et~al., 2020]{urstoger2020microstructure}
	Urst{\"o}ger, G., Kulachenko, A., Schennach, R., and Hirn, U. (2020).
	\newblock Microstructure and mechanical properties of free and restrained dried
	paper: a comprehensive investigation.
	\newblock {\em Cellulose}, 27(15):8567--8583.
	
	\bibitem[Vonk et~al., 2021]{vonk2021full}
	Vonk, N.~H., Geers, M. G.~D., and Hoefnagels, J. P.~M. (2021).
	\newblock Full-field hygro-expansion characterization of single softwood and
	hardwood pulp fibers.
	\newblock {\em Nordic Pulp \& Paper Research Journal}, 36(1):61--74.
	
	\bibitem[Vonk et~al., 2023a]{vonk2023res}
	Vonk, N.~H., Peerlings, R. H.~J., Geers, M. G.~D., and Hoefnagels, J. P.~M.
	(2023a).
	\newblock Effect of restrained versus free drying on hygro-expansion of
	hardwood and softwood fibers and paper handsheet.
	\newblock {\em Submitted to arXiv}.
	
	\bibitem[Vonk et~al., 2023b]{vonk2023bonds}
	Vonk, N.~H., Peerlings, R. H.~J., Geers, M. G.~D., and Hoefnagels, J. P.~M.
	(2023b).
	\newblock Full-field, quasi-3d hygroscopic characterization of paper
	inter-fiber bonds.
	\newblock {\em Submitted to arXiv}.
	
	\bibitem[Vonk et~al., 2023c]{vonk2023frc}
	Vonk, N.~H., Peerlings, R. H.~J., Geers, M. G.~D., and Hoefnagels, J. P.~M.
	(2023c).
	\newblock Re-understanding the in-plane hygro-expansion of freely and
	restrained dried paper handsheets.
	\newblock {\em Fundamental Pulp and Paper Physics Symposium (Accepted for
		publication)}.
	
	\bibitem[Vonk et~al., 2020]{vonk2020robust}
	Vonk, N.~H., Verschuur, N. A.~M., Peerlings, R. H.~J., Geers, M. G.~D., and
	Hoefnagels, J. P.~M. (2020).
	\newblock Robust and precise identification of the hygro-expansion of single
	fibers: a full-field fiber topography correlation approach.
	\newblock {\em Cellulose}, 27(12):6777--6792.
	
	\bibitem[Weise and Paulapuro, 1995]{weise1995changes}
	Weise, U. and Paulapuro, H. (1995).
	\newblock Changes of pulp fibre dimensions during drying.
	\newblock In {\em International Paper Physics Conference, Niagara-on-the-Lake,
		11.-14.9. 1995}, pages 121--124. Technical Section CPPA \& TAPPI.
	
	\bibitem[Wernersson et~al., 2014]{wernersson2014characterisations}
	Wernersson, E. L.~G., Borgefors, G., Borodulina, S., and Kulachenko, A. (2014).
	\newblock Characterisations of fibre networks in paper using micro computed
	tomography images.
	\newblock {\em Nordic Pulp \& Paper Research Journal}, 29(3):468--475.
	
\end{thebibliography}
\end{document}